\documentclass{ieeeaccess}

\usepackage{cite}
														  
\usepackage[colorlinks,urlcolor=blue]{hyperref}
\hypersetup{
    colorlinks=true,
    linkcolor=blue,
    citecolor=blue,
    urlcolor=blue,
    }

\usepackage{graphicx}

\usepackage{textcomp}
\def\BibTeX{{\rm B\kern-.05em{\sc i\kern-.025em b}\kern-.08em
    T\kern-.1667em\lower.7ex\hbox{E}\kern-.125emX}}
\usepackage[english]{babel}

\usepackage{subcaption}
\usepackage[labelsep=period]{caption}
\usepackage[exponent-product=\cdot]{siunitx}

\usepackage{float}

\usepackage{tabularx}
\usepackage{booktabs}

\usepackage[nolist,nohyperlinks]{acronym}

\usepackage[final]{microtype}

\usepackage[intlimits]{amsmath}

\usepackage[nameinlink,capitalize]{cleveref}

\Crefname{figure}{Fig.}{Figs.}
\Crefname{table}{Table}{Tables}
\DeclareCaptionFormat{figcapfont}{\figcapfont \textbf{\textcolor{accessblue}{#1#2}}\textsf{#3}}
\def\subfigcapfont{\rmfamily\fontencoding{T1}\fontseries{n}\fontsize{8}{9.6}\selectfont} 
\DeclareCaptionFormat{subfigcapfont}{\subfigcapfont #1#2#3}
\renewcommand{\thetable}{\Roman{table}}

\begin{document}
    \begin{acronym}
        \acro{cim}[CIM]{constant interaction or capacitance circuit model}
        \acro{csd}[CSD]{charge stability diagram}
        \acro{dc}[DC]{direct current}
        \acro{dqd}[DQD]{double quantum dot}
        \acro{idt}[IDT]{interdot transition}
        \acro{ldt}[LDT]{lead-to-dot transition}
        \acro{mps}[MPS]{matrix product state}
        \acro{pcb}[PCB]{printed circuit board}
        \acro{psd}[PSD]{power spectral density}
        \acro{qcad}[QCAD]{quantum computer-aided design}
        \acro{qtcad}[QTCAD]{quantum-technology computer-aided design}
        \acro{qd}[QD]{quantum dot}
        \acro{qubit}[qubit]{quantum bit}
        \acro{rf}[RF]{radio frequency}
        \acro{rtn}[RTN]{random telegraph noise}
        \acro{sd}[SD]{sensing dot}
        \acro{tcad}[TCAD]{technology computer-aided design}
        \acro{tct}[TCT]{total charge transition}
        \acro{tfcm}[TFCM]{Thomas-Fermi capacitive model}
        \acro{wkb}[WKB]{Wentzel-Kramers-Brillouin}
        \acro{2deg}[2DEG]{two-dimensional electron gas}
        \acro{seq}[SEQ]{Schrödinger equation}
        \acro{meq}[MEQ]{master equation}
        \acro{dft}[DFT]{density functional theory}
        \acro{gf}[GF]{Green function}
        \acro{mc}[MC]{Monte Carlo}
        \acro{hm}[HM]{Hubbard model}
        \acro{ml}[ML]{machine learning}
        \acro{lbmeq}[LBMEQ]{Lindblad master equation}
        \acro{brmeq}[BRMEQ]{Bloch-Redfield master equation}
        \acro{tbm}[TBM]{tight-binding model}
        \acro{peq}[PEQ]{Poisson equation}
        \acro{fem}[FEM]{finite-element method}
        \acro{ci}[CI]{configuration interaction}
        \acro{1vn}[1vN]{first-order von Neumann}
        \acro{2vn}[2vN]{second-order von Neumann}
        \acro{rtd}[RTD]{real-time diagrammatic}
    \end{acronym}

    \doi{-}
    
    \title{Simulation of Charge Stability Diagrams for Automated Tuning Solutions (SimCATS)}

    \author{\uppercase{Fabian~Hader}\authorrefmark{1},
            \uppercase{Sarah~Fleitmann}\authorrefmark{1},
            \uppercase{Jan~Vogelbruch}\authorrefmark{1},
            \uppercase{Lotte~Geck}\authorrefmark{1,2},
            \uppercase{and Stefan~van~Waasen}\authorrefmark{1,3}}
    \address[1]{Central Institute of Engineering, Electronics and Analytics
    ZEA-2 -- Electronic Systems, Forschungszentrum Jülich GmbH, 52425 Jülich,
    Germany}
    \address[2]{System Engineering for Quantum Computing, Faculty of Electrical Engineering and Information Technology, RWTH Aachen University, 52062 Aachen}
    \address[3]{Faculty of Engineering -- Communication Systems, University of Duisburg-Essen, 47057 Duisburg, Germany}

    \markboth
    {Hader \headeretal: Simulation of CSDs for Automated Tuning Solutions (SimCATS)}
    {Hader \headeretal: Simulation of CSDs for Automated Tuning Solutions (SimCATS)}

    \corresp{Corresponding author: Fabian Hader (email: f.hader@fz-juelich.de).}

    \begin{abstract}
     Quantum dots must be tuned precisely to provide a suitable basis for quantum computation. A scalable platform for quantum computing can only be achieved by fully automating the tuning process. One crucial step is to trap the appropriate number of electrons in the quantum dots, typically accomplished by analyzing \acp{csd}. Training and testing automation algorithms require large amounts of data, which can be either measured and manually labeled in an experiment or simulated. This article introduces a new approach to the realistic simulation of such measurements. Our flexible framework enables the simulation of ideal \ac{csd} data complemented with appropriate sensor responses and distortions. We suggest using this simulation to benchmark published algorithms. Also, we encourage the extension by custom models and parameter sets to drive the development of robust, technology-independent algorithms. Code is available at \href{https://github.com/f-hader/SimCATS}{https://github.com/f-hader/SimCATS}.
    \end{abstract}
    

    \titlepgskip=-15pt
    \maketitle

    \section{Introduction}
    \label{sec:introduction}

    \PARstart{Q}{uantum} dot tuning automation is a crucial step to enable a scalable platform for quantum computation. Two essential steps are the isolation of electrons in \acp{qd}, hereafter referred to as dot regime tuning, and the adjustment of an appropriate number of electrons, referred to as charge state tuning below. One can observe the charge and spin states in gate-defined \acp{qd} by the conductance change of a nearby electrostatically coupled sensor dot. The development of tuning algorithms based on machine learning and classical algorithms, the assessment of the quality of a solution, and the comparison of different approaches benefit from publicly available datasets. Especially for the latter two purposes, this is even a prerequisite to enable fair comparability. Simulations can generate the required number of datasets along with the corresponding ground truth data. Therefore, we propose a generic framework for the simulation of \acp{csd} that combines the necessary functionalities to mimic experimental data. To simulate the ideal \ac{csd}\footnote{In this context, ideal refers to simulated undisturbed (ground truth) data.}, we introduce a geometric model, which does not require knowledge of the physical device parameters. Instead, it allows the reconstruction of measurement data merely based on parameters describing the geometry observed in previously recorded data. 
    
    This paper is organized as follows: First, we comprehensively introduce the different aspects of quantum simulation (\cref{sec:background}). Then, we describe our simulation model (\cref{sec:model}) consisting of our geometric model for the \ac{dqd} occupation (\cref{ssec:model_occ}), a model for the sensor response (\cref{ssec:model_sensor}), and distortions (\cref{ssec:model_noise}). Next, we depict the extraction of parameters from measured data (\cref{sec:parameter_extraction}), evaluate the quality of the simulated data (\cref{sec:evaluation}), and, finally, summarize our work and draw a conclusion comprising suitable application fields and prospective improvements (\cref{sec:conclusion}).

    \section{Background}
    \label{sec:background}

    Simulating quantum mechanical processes is a manifold and complex task, especially for many-body systems and their corresponding Hamiltonian due to the exponential scaling of the required resources with the number of particles in the system and the large number of environmental degrees of freedom. Several theories and models that describe different quantum mechanical processes include the \ac{seq} and \acp{meq} for closed and open quantum systems \cite{grabert_1992, nazarov_1993, konig_1996, timm_2008}, Feynman path integrals \cite{feynman_1963}, the \ac{dft} \cite{singh_1994, siano_2004, pokorny_2018}, mean-field theories \cite{hepp_1974, ginibre_1979, erdos_2001}, nonequilibrium \acp{gf} \cite{bulka_2004, sztenkiel_2007, zagoskin_2014}, and the random/scattering matrix theory \cite{brody_1981, newton_1982, beenakker_1997, potz_2008}. Besides the time dynamics of the quantum states (described by \ac{seq} and \ac{meq}), the ground state (the eigenvector of the Hamiltonian with the smallest eigenvalue) is of fundamental interest and corresponds to the state when the system is at zero temperature. Principally, classical computers or quantum devices can be used for the simulation task \cite{feynman_1982}. 

    Quantum simulators use a quantum system to model a Hamiltonian. They can be digital, i.e., they use a quantum computer with qubits and sequentially applied gates, or analog, i.e., the specially designed system emulates the Hamiltonian. The first approach is more general but requires thousands of highly controllable qubits. The second incorporates no gates and is easier to control but less versatile, on the other hand. \cite{georgescu_2014} described a comprehensive overview of the proposed systems and potential application fields. Moreover, \ac{qd} systems \cite{vanderwiel_2002, hanson_2007} play an exciting role in this field recently, e.g., in simulating the low-temperature \ac{hm} \cite{barthelemy_2013}, emulating Fermi-Hubbard models \cite{byrnes_2008, hensgens_2017, hensgens_2018, wang_2022}, demonstrating Nagaoka ferromagnetism \cite{dehollain_2020}, or simulating the antiferromagnetic Heisenberg chain \cite{vandiepen_2021, vandiepen_2021a}. 
    
    When using classical computer systems to simulate higher particle systems, numerical approximation approaches must be used. Here, the challenge is to find the balance between exactness, computation cost, applicability to the problem, and validity of results. Methods proposed for this task comprise quantum \ac{mc} approaches \cite{suzuki_1993, siano_2004, luitz_2010, pokorny_2018}, many-body perturbation theories \cite{dreizler_1990, fetter_2003}, multi-configuration time-dependent Hartree \cite{meyer_1990, manthe_1992, beck_2000}, hierarchical equations of motion \cite{tanimura_1989, tanimura_1990, tanimura_2006}, machine learning \cite{carleo_2017, hartmann_2019, schutt_2019, hermann_2020, manzhos_2020}, and tensor networks \cite{vidal_2004, hong_2022,orus_2019}. The latter include the numerical renormalization group \cite{wilson_1975, yoshioka_2000, oguri_2004, bulla_2008, oguri_2013}, the \ac{mps} \cite{fannes_1992, vidal_2003, verstraete_2008, cirac_2021} as a particular case for 1D systems, and the density matrix renormalization group \cite{white_1992, white_1993, takasaki_1999, schollwock_2005, schollwock_2011a} as a variational algorithm in the set of MPS \cite{ostlund_1995, dukelsky_1998, verstraete_2004}. However, perfect or even good models and approximations are not always available or require too much processing capacity, even for today's supercomputers.
    
    Simulating the transport of semiconductor \ac{qd} systems on classical computers is a demanding task that incorporates phenomena like Coulomb blockade \cite{beenakker_1991, grabert_1992, livermore_1996}, Pauli spin blockade \cite{ono_2002}, or sensor dot response \cite{elzerman_2003}, for example. Specific models ease the calculation of \acp{csd} \cite{hofmann_1995} that are fundamental for spin-based quantum computation regarding qubit manipulation and information readout.
    
    The \ac{cim} \cite{kouwenhoven_1997, vanderwiel_2002, hanson_2007, schroer_2007} describes the electronic states of \acp{qd} and parametrizes the onsite and intersite Coulomb interaction as a network of capacitors, leaking capacitors, and resistors between dots, gates, and leads. It explains many aspects of experiments satisfactorily, but several quantum effects deform the modeled \acp{csd}, sometimes even substantially. 
    Theories capable of considering both classical and quantum effects help to understand the quantum aspects of \acp{csd} and improve their usefulness. The \ac{tfcm} of \cite{kalantre_2019} uses the Thomas-Fermi approximation \cite{march_1983} to calculate the electron density of the islands and derive an estimate of inverse capacitive elements for a capacitance model in a given potential profile. It also models the electron transport using a Markov chain among charge states incorporating single electron tunneling rates between islands or to contacts calculated under the \ac{wkb} tunnel probability \cite{merzbacher_1998a}. The quantum-mechanical two-level model \cite{dicarlo_2004, petta_2004, hatano_2005, huttel_2005, pioro-ladriere_2005} derives the tunneling of a single electron between two dots from the probability crossover of the two eigenstates and \cite{zhang_2006} investigated its influence on the \ac{csd}. The first interest in applying \acp{hm} to \ac{qd} systems appears in the field of collective Coulomb blockade \cite{stafford_1994, kotlyar_1998}. An effective charge-spin model for \acp{qd} \cite{jefferson_1996} based on a lattice description equivalent to a single-band \ac{hm} incorporated higher-order perturbation theory and \ac{wkb} approximation. Nevertheless, \cite{gaudreau_2006, korkusinski_2007} fundamentally demonstrated the capability of \acp{hm} to describe \acp{csd} for triple-quantum-dot systems. Later, \cite{yang_2011} introduced a generalized \ac{hm} as the quantum generalization of the classical \ac{cim}, including quantum effects such as spin exchange, pair-hopping, and occupation-modulated hopping. Experiments on silicon systems \cite{dassarma_2011} quantitatively confirmed the model's applicability, and the effects of the involved quantum parameters on \acp{csd} have been discussed in-depth \cite{wang_2011b}. To calculate \acp{csd} under a lead bias, \cite{rassekh_2022} used Fermi’s golden rule to obtain the transition rates, extracted the probabilities of the states, and finally calculated the current. Nevertheless, the generalized \ac{hm} concentrates on the electronic interaction in the quantum-dot system itself and neglects environmental factors \cite{storcz_2005, stavrou_2005, witzel_2006, taylor_2007, gimenez_2009, nguyen_2011a}.
    
    The models often find their application in \ac{ml} approaches for automated measurement and tuning of \acp{qd}. \cite{lennon_2019} used the \ac{cim} to simulate current maps of single \acp{qd} and learned an algorithm that automatically chooses the most informative subsequent measurements. To determine the system’s virtual voltages, \cite{oakes_2021} applied \ac{cim} simulated and experimental \ac{csd} data of a 2x2 \ac{qd} array to train and validate regression models for the extraction of the gradients from a Hough transformation \cite{oakes_2021}. A purely theoretical approach to \ac{cim} simulated data \cite{krause_2022} tries to find the most probable convex polytope of Coulomb diamonds in \ac{qd} measurements by learning a device model using regularized maximum likelihood estimation and one-dimensional raster scans (rays) only. \cite{li_2023} studied the effects of involved quantum parameters on \acp{csd} of a serial triple \ac{qd} and confirmed their global features by the similarity between transport measurements and \ac{cim}-based simulations. To detect charge states, \cite{darulova_2020a} evaluated the prediction accuracy of several machine learning models trained on simulated and experimental data. The simulated data are generated from \ac{cim} or taken from the Qflow-lite dataset \cite{zwolak_2018}, both improved with five different noise types added. The Qflow-lite dataset is based on the \ac{tfcm} realization of \cite{kalantre_2019}, used to develop deep and convolutional neural networks to tune \ac{qd} arrays automatically to a double-dot configuration. The dataset intends to provide a reliable dataset of simulated device state (state labels), current, charges, and charge sensor response versus the gate voltages. It extends the \ac{tfcm} by a charge sensor response. Qflow 2.0 \cite{ziegler_2022} constitutes a further refinement of the dataset by adding synthetic noise characteristic of \ac{qd} devices. It is part of a framework for robust automated tuning that uses convolutional neural networks for device state estimation and data quality control.
    
    However, implementations of classical simulators for quantum systems are manifold and numerous. Broader application fields are covered, e.g., by QuTiP \cite{johansson_2012, johansson_2013}, QuantumATK \cite{smidstrup_2019}, and Kwant \cite{groth_2014}, and comprise a set of basic solvers. QuTiP offers solvers for the time evolution of open quantum systems comprising \ac{lbmeq} and \ac{mc} solvers, routines for the \ac{brmeq}, periodic systems using the Floquet formalism, and stochastic solvers. Differently, the Kwant Python package provides numerical calculations on \acp{tbm} with a strong focus on quantum transport. Currently, the Coulomb blockade is not supported directly. QuantumATK offers a fully integrated Python/C++ platform of electronic and atomic-scale modeling tools for electronic structure calculations (via DFT, semi-empirical \ac{tbm} Hamiltonians, classical and \ac{ml} force fields) and electron transport simulations (via \ac{gf}) supporting \ac{csd} plots.
    
    Another group of simulators concentrates on the simulation and design of semiconductor-based information devices. The nextnano/nextnano++ \cite{birner_2007} 3D simulator computes electron transport (via \ac{seq}, \ac{peq}, and current equations) and \acp{csd} for arbitrary designs. The \ac{qcad} software \cite{gao_2013a, nielsen_2013} primarily designs and models silicon multi-\acp{qd} developed for qubits. It implements a \ac{fem} based tool that contains nonlinear \ac{peq}, effective mass \ac{seq}, and \ac{ci} solvers. Currently, magnetic fields and direct \ac{csd} outputs are not supported. QmeQ \cite{kirsanskas_2017} focuses on the numerical modeling of stationary-state transport through \ac{qd} devices (via Pauli \ac{meq}, \ac{lbmeq}, Redfield \ac{meq}, and \ac{1vn} approaches) with strong electron-electron interactions. It also computes co- and pair-tunneling (via \ac{2vn} and \ac{rtd}) and broadening of \ac{qd} states (via \ac{rtd}), leading to \acp{csd} that include quantum effects. The integrated device simulator for quantum bit design based on impulse \ac{tcad} \cite{ikegami_2019,asai_2021a} offers quantum bit output, quantum transport, and qubit operations. The computation pipeline consists of \ac{qd} potential calculation (via \ac{seq} and \ac{peq} coupled with semi-classical drift-diffusion), \ac{qd} capacitance calculation (via fictitious charge change), micromagnetic simulations (via Ampere's law solved by finite-volume method), and single-electron quantum transport (via \ac{seq} with magnetic field). Even more focused, \ac{qtcad} \cite{beaudoin_2022} implements a \ac{fem} simulator to predict the performance of spin-qubit devices. Incorporated methods include nonlinear \ac{peq}, \ac{seq}, \ac{meq}, and many-body solvers. Quantum transport calculations in the sequential tunneling regime enable the treatment of Coulomb blockade and the calculation of \acp{csd}.
    
    Finally, some simulators only focus on \ac{csd} simulation. SIMON \cite{wasshuber_1997a} simulates single-electron tunnel devices and circuits using the \ac{cim} and the \ac{mc} method for \ac{meq} to implement tunnel junctions. Although designed as a Python-based framework for the tuning and calibration of \acp{qd} and spin qubits, QTT \cite{eendebak_2023} offers \ac{cim}-based \ac{csd} simulation functions. From the system parameters provided, they set up the Hamiltonian, compute the eigenvalues, determine occupation numbers, and finally derive the \ac{csd}. Additionally to the \ac{cim}-based simulation, the QuDiPy project \cite{qudipy} implements the \ac{hm} of \cite{dassarma_2011} to generate simulated \acp{csd}.

    \section{Simulation model}
    \label{sec:model}
    
\begin{figure*}[ht]
        \centering
        \includegraphics[width=0.8\linewidth]{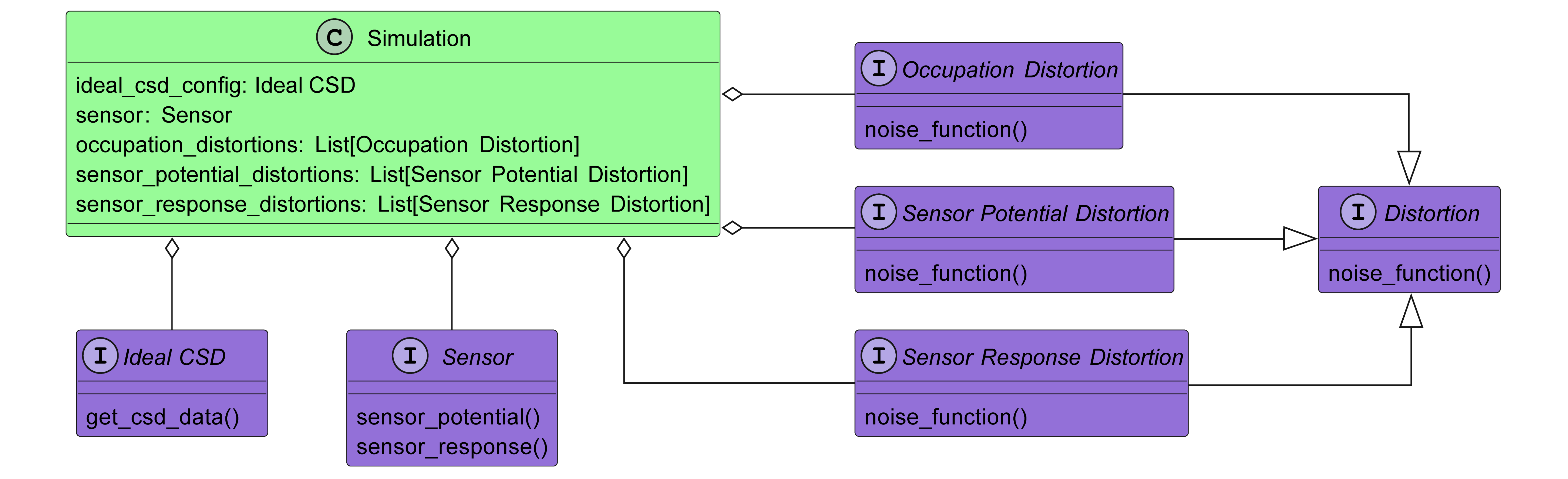}
        \caption{Interfaces of the simulation class of our simulation framework. The Python package includes standard implementations.}
        \label{fig:simcats_interfaces}
    \end{figure*}

    Our simulation model combines the simulation of the occupation probabilities (\cref{ssec:model_occ}), the sensor response (\cref{ssec:model_sensor}), and several types of distortions (\cref{ssec:model_noise}) into a single framework to provide a comprehensive simulation of \acp{csd}. We conceptualized the handling of the framework to mimic the experiment as well as possible. In particular, we focus on integrating as much flexibility as possible to support different types of dot regimes and charge state tuning experiments. Therefore, the framework enables users
    \begin{itemize}
        \item to perform 2D and 1D measurements,
        \item to measure in different directions with consideration of the time dependence of certain distortion types,
        \item to switch to different sensor configurations, such as for multi-sensor samples, and
        \item to switch off distortions individually.
    \end{itemize} 
    All parts of the simulation are interchangeable and defined via simple interfaces, as shown in \cref{fig:simcats_interfaces}.

    \subsection{Dot occupation model}
    \label{ssec:model_occ}
    
\begin{figure}[ht]
        \centering
        \includegraphics[width=\linewidth]{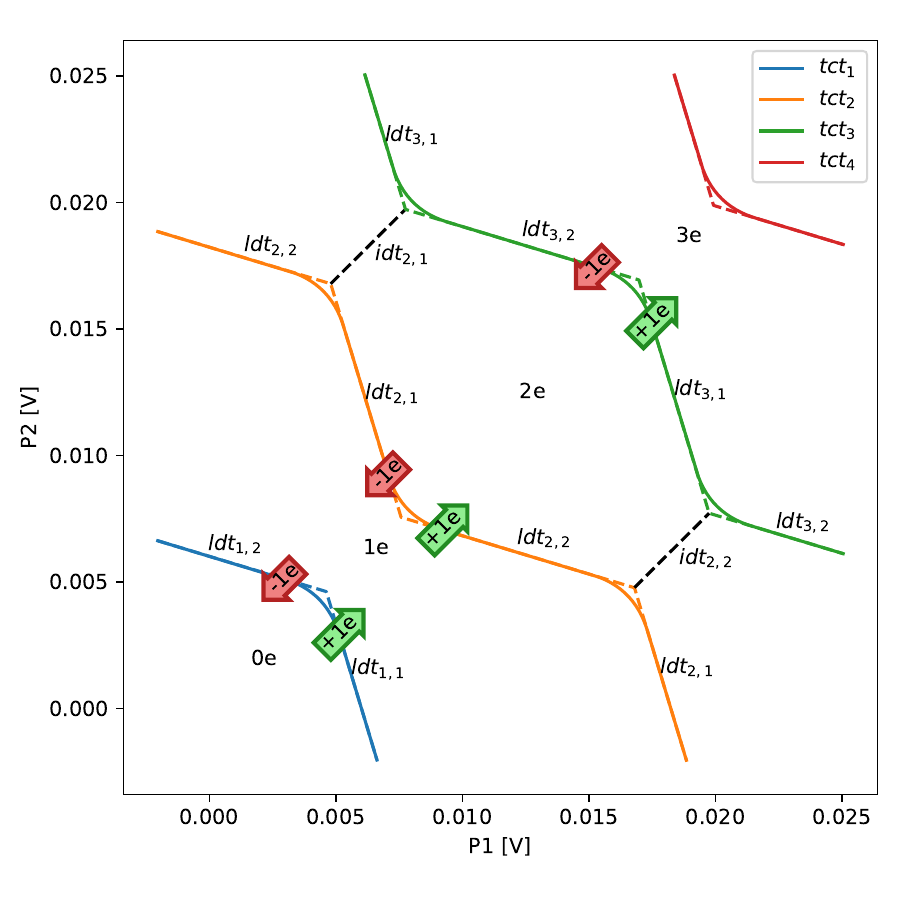}
        \caption{Representation of the separation between regions with fixed numbers of electrons using \acp{tct}. The solid lines represent \acp{tct} affected by interdot tunnel coupling, and the dashed lines indicate the \acsp{ldt} without tunnel coupling.}
        \label{fig:2D_tct_example}
    \end{figure}

    The electron occupation of the dots is the underlying basis for the \ac{csd} simulation. Our occupation model is purely geometric and, in contrast to available physical models, provides the flexibility and simplicity to support the different honeycomb shapes we observe in \ac{dqd} measurements. The fundamental idea is to describe a \ac{csd} as a series of \acp{tct} representing the borders between regions containing a fixed number of electrons in the system. \cref{fig:2D_tct_example} illustrates an example of this representation, where \(tct_i, i=1,...,n\) separates the regions containing \(i-1\) and \(i\) electrons.
    
    The orientation of the \acp{ldt}\footnote{A \ac{ldt} is a transition where an electron tunnels in (or out) of the dot system from (or to) the leads.} in the two-dimensional voltage space depends on the capacitive coupling of the gates. \(ldt_{i,j}\) (associated to \(tct_i\)) is primarily affected by plunger gate\footnote{Gates used to primarily control the potential of a dot.} \(P_{j}\) of \(qd_{j}\) or, in case of using virtual gates\footnote{Virtual gates are a linear combination of multiple physical gates used to compensate for capacitive coupling and influence only one parameter of the system}, even exclusively by virtual plunger gate \(P^{v}_{j}\). For the representation in \cref{fig:2D_tct_example}, \(ldt_{i,1}\) has a slope in the interval \([-\infty, -1)\) (\(-\infty\) for virtual gates) and \(ldt_{i,2}\) in the interval \((-1, 0]\) (0 for virtual gates). To mathematically represent the slopes unambiguously, we propose a parametric representation of the \acp{tct} in a voltage space \((V'_{P1}, V'_{P2})\) originating from \((V_{P1}, V_{P2})\) by an affine transformation corresponding to a 45\(^{\circ}\) rotation. This results in slope intervals of \([-1, 0)\) for \(ldt_{i,1}\) and \((0, 1]\) for \(ldt_{i,2}\).
    Depending on the interdot tunnel coupling, the \ac{tct} exhibits curves at triple points, where \(ldt_{i,1}\), \(ldt_{i,2}\), and an \ac{idt}\footnote{An \ac{idt} describes the tunneling of an electron from one dot to another.} intersect. Thus, composing a \ac{tct} of linear parts and B\'{e}zier curves ensures twice continuous differentiability.
    
    The following parameters define \(tct_i\):
    \begin{enumerate}
        \item \(s_{i,j}\): \(V'_{P1}\)-intercept of \(ldt_{i,j}, (j=1,2)\),
        \item \(m_{i,j}\): slope of \(ldt_{i,j}, (j=1,2)\),
        \item \(b_{i,2}\): B\'{e}zier anchor on \(ldt_{i,2}\) defining the starting point of the curved part of the first triple point of \(tct_i\),
        \item \(b_{i,1}\): B\'{e}zier anchor on \(ldt_{i,1}\) defining the end point of the curved part of the first triple point of \(tct_i\).
    \end{enumerate}
    Using only this set of parameters, enables the complete construction of a \ac{tct} out of repetitions of the linear parts, the B\'{e}zier curve, and its 180\(^{\circ}\) rotation. 
    Furthermore, the intersection of the two \acp{ldt} constitutes the required center B\'{e}zier anchor \(b_{i}\):
    \begin{align}
        \label{eq:center_bezier_anchor}
        P1'_{b_{i}} & = \frac{P2'_{b_{i,2}} - P2'_{b_{i,1}} + P1'_{b_{i,1}} \cdot m_{i,1} - P1'_{b_{i,2}} \cdot m_{i,2}}{m_{i,1} - m_{i,2}} \\
        P2'_{b_{i}} & = P2'_{b_{i,2}} + m_{i,2} \cdot (P1'_{b_{i}} - P1'_{b_{i,2}}) \,.
    \end{align}
    The \(V'_{P1}\)-intercept of only the linear part is
    \begin{equation}
    	\label{eq:length_linear}
    	l_{i,j} = s_{i,j} - 2 \cdot |(P1'_{b_{i,j}}-P1'_{b_{i}})| \,.
    \end{equation}
    
    Depending on the identifier \(i\) of the \ac{tct}, the number of B\'{e}zier curves and triple points is limited to
    \begin{equation}
        n_{t} = 2 \cdot i - 1 \,.
    \end{equation}
    \begin{figure*}[ht]
        \centering
        \includegraphics[width=0.7\linewidth]{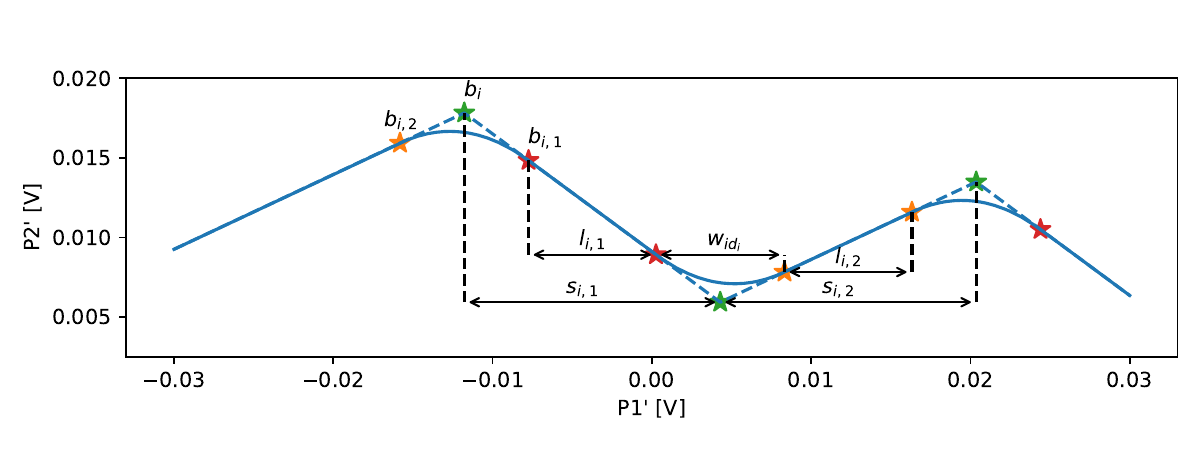}
        \caption{\ac{tct} description in 1D (rotated). The blue dashed line represents the \ac{tct} without tunnel coupling and the blue solid line with tunnel coupling.}
        \label{fig:1D_tct_description}
    \end{figure*}
    
    Existing \acp{tct} allow the calculation of the electron occupation. In the area between \(tct_i\) and \(tct_{i+1}\), a total of \(i\) electrons is in the system. Their distribution to the two \acp{qd} \(qd_{1}\) and \(qd_{2}\) is determined as follows:
    \begin{enumerate}
        \item The connecting vector from the triple point of \(tct_i\) to the opposite triple point of \(tct_{i+1}\) represents the interdot transition \(idt_{i,k}, k \in \{1, ..., i\}\).
        \item Across each \(idt_{i,k}\), a sigmoid function orthogonal to it approximates the Fermi distribution. 
        \item The superposition\footnote{Here, superposition denotes mathematical function combination.} of all sigmoid functions represents the electron occupation of \(qd_{1}\).
        \item The difference between the number of total charges \(i\) and the occupation of \(qd_{1}\) results in the occupation of \(qd_{2}\) .
    \end{enumerate}  
    
    \subsection{Sensor model}
    \label{ssec:model_sensor} 
    
\begin{figure}[ht]
        \centering
        \includegraphics[]{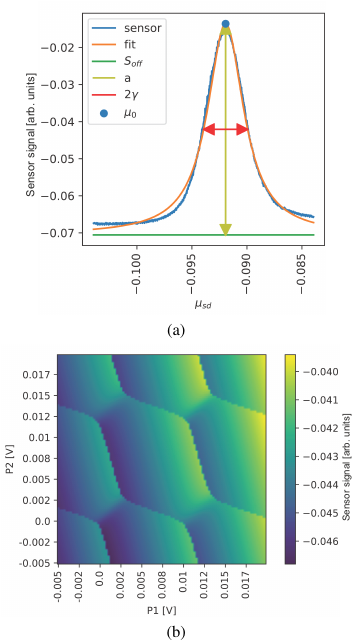}
    	\caption{Simulation of the sensor response. (a) Measured sensor response (blue) with Lorentzian fit (orange) (b) Example of the sensor response simulated on the left flank of the peak, resulting in rising value shifts with rising voltages in the \ac{csd}.}
        \label{fig:sensor_response}
    \end{figure}

    We calculate the sensor response at each point in the \ac{csd} using the simulated occupation information and the sensor characteristic. \\
    Besides the required capacitive coupling of the \ac{sd} to the \ac{dqd},  the \ac{sd} also cross-couples to the plunger gates of the \ac{dqd}. In \acp{csd}, the first enables the observation of electron occupation changes as edges, whereas the second appears as undesired value shifts inside the honeycombs\footnote{The undesired coupling between the \ac{dqd} plunger gates and the \ac{sd} can be compensated by using virtual gates.}. The simulation should incorporate both. \\
    We propose the following model for the sensor response $S$ \cite{fleitmann_2023}:
    \begin{equation}
    	\label{equ:lorentzian}
    	\begin{gathered}
    		\mu_{sd} = \sum_{j=1}^{2} \left[ \alpha_{j} \cdot N_{j} + \beta_{j} \cdot V_{P_{j}} \right] + \mu_{sd,0} \\
    		S = S_{off} + a \cdot \frac{\gamma^2}{\gamma^2 + (\mu_{sd} - \mu_0)^2},
    	\end{gathered}
    \end{equation}
    where $j$ is the index of the corresponding plunger gate.
    In this model, $\mu_{sd}$ represents the electrochemical potential of the \ac{sd} influenced by the number of electrons $N_{j}$ in the dots and the voltages applied to the plunger gates $V_{P_{j}}$ together with their corresponding lever arms $\alpha$ and $\beta$. Effectively, $\alpha$ influences the sharpness of the edges and $\beta$ the drifts within the honeycombs. Moreover, both effects are counteractive: $\alpha$ is negative, whereas $\beta$ is positive. Furthermore, the initial potential $\mu_{sd,0}$ of the SD adds to the potential. \\
    A simplified Lorentzian \cite{kouwenhoven_1997} approximates $S$ (see \cref{fig:sensor_response}a). $\gamma$ defines its width, and $\mu_0$ the potential at the peak. As linear filters transform the sensor response in the experimental setup, a scaling factor $a$ and an offset $S_{off}$ parametrize the Lorentzian.
    \cref{fig:sensor_response}b shows an example of a simulated \ac{csd} that includes the cross-coupling effects. \\
    
    \subsection{Distortions Model}
    \label{ssec:model_noise}
    The simulation of realistic \acp{csd} requires the consideration of occurring distortions \cite{darulova_2020a, ziegler_2022, fleitmann_2023}. In the following, we define identified distortion phases, assign collected distortion types to these, and describe their sources, simulation, and the required parameters as they typically appear in measurement setups similar to \cite{hader_paper, botzem_tuning_2018}. Additionally, we assume samples and their layout to be good enough for scalability (e.g. no spurious \ac{qd} under the barrier gates or intensively moving \acp{qd}), as this is a prerequisite to build a functional quantum computer. Under the assumption that the measurement is performed slow enough, effects like latching can be neglected here.
    
    \subsubsection{Distortion Categories}
    We propose to assign the \ac{csd} distortions to three categories based on their impact point in the signal path:
    \begin{enumerate}
        \item Occupation distortions,
        \item \Ac{sd} potential distortions, and
        \item \Ac{sd} response distortions.
    \end{enumerate}
    Distortions of the first category alter the simulated occupations of the \ac{dqd}. The category includes dot jumps (\cref{sssec:dot_jumps_desc}) and temperature broadening (\cref{sssec:temperature_broadening}). \\
    \Ac{sd} potential distortions comprise \ac{rtn} (\cref{sssec:rtn_description}) and pink noise (\cref{sssec:pink_noise}). It is crucial to differentiate these from undesired effects on the nonlinear \ac{sd} response, primarily white noise (\cref{sssec:white_noise}) and \ac{rtn} (\cref{sssec:rtn_description}).

    \subsubsection{Dot Jumps}
    \label{sssec:dot_jumps_desc}
    Dot jumps originate from deterministic charge-trapping effects on the \acp{qd} caused by fabrication-related imperfections. They become visible as displacements inside the occupation structure of a \ac{csd} \cite{ziegler_2022}. We simulate them by shifting a block of columns horizontally, like in \cref{fig:noise_examples}c, or a block of rows vertically.\\
    A geometric distribution of the jump extension simulates their occurrence, whereas a Poisson distribution of the jump amplitude determines their intensity \cite{ziegler_2022}.
    
    \subsubsection{Occupation Transition Blurring}
    \label{sssec:temperature_broadening}
    The thermal occupation of states in the lead reservoir leads to a broadening of the \acp{ldt}, which follows the Fermi-Dirac distribution under the assumption that the density of states is constant \cite{maradan_gaas_2014}. We simulate this effect by applying a one-dimensional Fermi-Dirac filter kernel along the measurement direction\footnote{For the presented results the Fermi-Dirac filter was still approximated by a Gaussian filter kernel.}.
    
    \subsubsection{Pink Noise}
    \label{sssec:pink_noise}
    Pink noise, also known as $1/f$ or flicker noise, is observed in most electronic devices and results from the internal heterogeneity of electronic components, such as oxide traps or lattice dislocations \cite{electronicNoise}. Its \ac{psd} is inversely proportional to the frequency and emerges as stripes in line-wise measured \acp{csd}.\\
    The generation of pink noise is described in \cite{pink_noise} and implemented in the Python module \texttt{colorednoise} \cite{colorednoise}. 
    In our simulation, pink noise is applied to the sensor potential, increasing its visibility in regions with high gradients due to the nonlinear sensor response. The latter corresponds to our observations in experimental data \cite{hader_paper}. 
    
    \subsubsection{Random Telegraph Noise}
    \label{sssec:rtn_description}
\begin{figure*}[ht]
        \centering
        \includegraphics[]{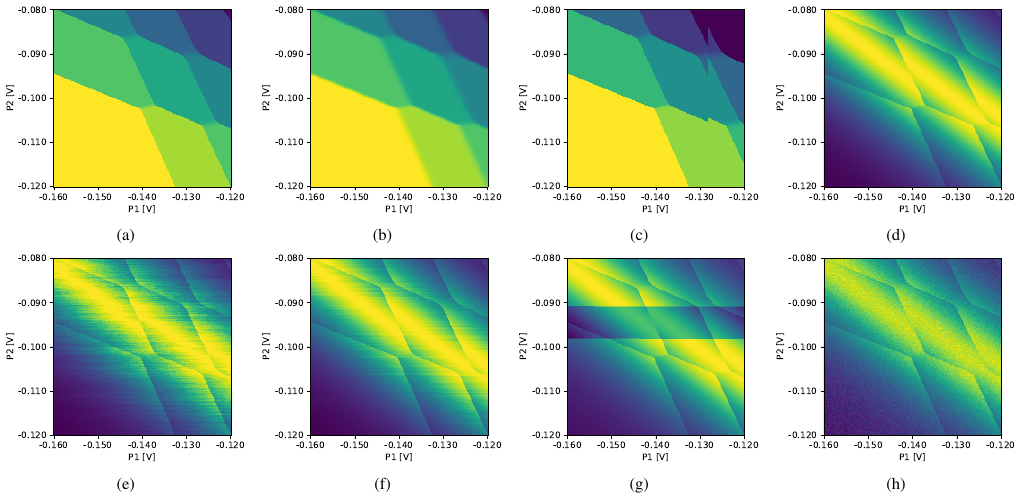}
        \caption{Examples of the simulated sensor response and distortions. Distortions affecting the sensor (cat. 2 and 3) are visualized in combination with a sensor response. (a) Ideal \ac{csd}, (b) ideal \ac{csd} with occupation transition blurring (cat. 1), (c) ideal \ac{csd} with dot jumps (cat. 1), (d) ideal \ac{csd} with sensor response, (e) ideal \ac{csd} with sensor response and pink noise (cat. 2), (f) ideal \ac{csd} with sensor response and \ac{rtn} (cat. 2), (g) ideal \ac{csd} with sensor response and \ac{rtn} (cat. 3), and (h) ideal \ac{csd} with sensor response and white noise (cat. 3).}
        \label{fig:noise_examples}
    \end{figure*}

    \Ac{rtn} or burst noise randomly switches between two or multiple discrete levels \cite{electronicNoise}. This effect results from a time-dependent random capture/emission process of charge carriers caused by oxide traps \cite{rtn}. Its \ac{psd} is proportional to $1/f^2$. In line-wise measured \acp{csd}, \ac{rtn} is visible as stripes with a well-defined beginning and ending (see \cref{fig:noise_examples}f).
    \\We simulate the occurrence of bursts using a geometric distribution for their extension and a normal distribution for their amplitude \cite{ziegler_2022}.\\
    Like pink noise, \ac{rtn} usually appears as noise on the sensor potential. However, the \acp{csd} also contain jumps that affect the sensor response. Therefore, we propose to include \ac{rtn} additionally in distortion category 3.
    
    \subsubsection{White Noise}
    \label{sssec:white_noise}
    In the system under consideration, white noise, having a constant \ac{psd} characteristic, originates from thermal \cite{thermal_noise} and shot noise \cite{shot_noise}. Thermal noise \cite{thermal_noise} is caused by the thermal agitation of charge carriers in an electrical conductor, whereas shot noise depends on the discrete charges in the current flow and does not relate to a system's operating temperature. \\
    The amplitude distribution is nearly Gaussian for thermal noise but Poissonian for shot noise. However, as a normal distribution can approximate the Poisson distribution, the simulation combines both noise types into one Gaussian distribution with standard deviation $\sigma_{w}$. Additionally, we assume that they solely accrue after the sensor, as their dominating parts result from the amplification of the sensor signal.
    
    \section{Parameter extraction}
    \label{sec:parameter_extraction}
    We exemplarily extracted parameters for the simulation of \acp{csd} from a GaAs sample\footnote{The determined parameters are provided with the simulation software (\texttt{default\_{}configs["GaAs\_{}v1"]}) \cite{simcats_github}.}, which is similar to \cite{volk_loading_2019}. Our extraction approach is independent of the sample used.
    
    \subsection{Extraction of occupation data parameters}
    Parameters describing the structure of \acp{csd} can be extracted directly from previously recorded measurements. \\
    Each considered \ac{tct} requires the parameters described in \cref{ssec:model_occ}. We extracted them by manual labeling. It is not necessary to save the parameters for all \acp{tct}. Instead, it is possible to define a transformation rule that generates the next \ac{tct} from a previous one, e.g. by shifting the \ac{tct} and adjusting \(w_{id}\) based on observed relations.

    \subsection{Extraction of sensor parameters}
    \label{ssec:parameters_sensor_response}
    In our case, the measured sensor scans describe the sensor response as a function of the voltage applied to the plunger gate of the corresponding \ac{sd}. Additionally, we utilize the proportionality of the electrochemical sensor potential to the \ac{sd} plunger gate in the following.\\
    Fitting the Lorentzian to the experimental sensor scan determines the parameters $S_{off}$, $a$, $\gamma$, and $\mu_0$ in the sensor model (\cref{ssec:model_sensor}).
    For the determination of $\alpha_j$ and $\beta_j$, we restrict the following analysis to areas in the \ac{csd} with an overall rising value $S(V_{P_1},V_{P_2})$, corresponding to the left side in the fit $S(\mu_{sd})$, due to the irreversible uniqueness of the Lorentzian. Then, we can use the inverse fit function $\mu_{sd}(S)$ to estimate the electrochemical sensor potential $\mu_{sd}(V_{P_1},V_{P_2})$. Inside honeycomb regions, we can determine
    \begin{equation}
        \beta_j=\left(\frac{\partial}{\partial V_{P_j}} \mu_{sd} \right)_{N_1,N_2=const} .
    \end{equation}

    Finally, inside the corresponding lead transition areas of the two dots, we calculate
    
    \begin{align}
        \alpha_i & =\left(\frac{\Delta (\mu_{sd}(V_{P_i})- \beta_i \cdot V_{P_i})}{\Delta N}\right)_{N_i \neq const, N_{j \neq i}=const}\textrm{.}
    \end{align}

    \subsection{Extraction of distortion parameters}
    We determine the distortion parameters from different scans. For white noise and \ac{rtn} of category 3 we use measured \acp{csd}; for pink noise and \ac{rtn} of category 2, we use sensor scans. However, we characterize dot jumps manually in measured \acp{csd}.
    
    \subsubsection{Dot Jumps}
    There is no method for detecting dot jumps automatically yet. Therefore, we extract the amplitude and extension of the jumps manually. However, as no return is visible in our \acp{csd}, only the intensity of the jumps can be determined. Thus, we assume the extension to be larger than the measured voltage space of the experimental \acp{csd}. Moreover, we extract the parameters for the two swept gates independently, as they might differ depending on the sample.

    \subsubsection{Pink Noise}
    \label{sssec:pink_parameters}
    The intensity of pink noise in the sensor potential can be determined using the \ac{psd}. For this purpose, we examine two-dimensional sensor scans with a high resolution on the abscissa and a low resolution on the ordinate \cite{hader_paper}.\\ 
    However, the sensor potential has to be estimated first. Therefore, for every measured row, a sum of Lorentzians is fitted. If successful\footnote{evaluated manually, cf. \cite{hader_paper}} for each measured gate voltage, we determine the corresponding sensor potential by computing the inverted Lorentzian of the measurement value.  \\
    Now, the \ac{psd} of the estimated potential can be determined for every row and averaged over different rows to get a better approximation. To obtain the intensities of the white and pink noise parts, we use the fit from \cref{sssec:white_parameters}, because the calculated electrochemical potential includes the white noise from the sensor response. Nevertheless, we extracted the white noise parameters used for the simulation directly from the sensor response.\\
    Tests of our estimation method with simulated data indicate that it usually overestimates the noise by a fixed factor that depends on the Coulomb oscillation characteristics of the sample.
    
    \subsubsection{Random Telegraph Noise}
    Currently, the automatic detection of \ac{rtn} and the separation from the pink noise in \acp{csd} is problematic. Therefore, we manually investigate sensor scans for \ac{rtn} of category 2 and determine the bursts’ extensions directly and the amplitudes from the jump in the calculated electrochemical sensor potential. Then, the mean of the extensions and the mean and empirical standard deviation of the amplitudes constitute the \ac{rtn} parameters. Translating the bursts' extension into the \ac{csd} domain during the simulation requires considering the measurement time per voltage due to the time-dependent stochastic nature of \ac{rtn}. We use the median of our \ac{csd} measurement time as the default value.\\
    For \ac{rtn} of category 3, we manually extract the extension and amplitude parameters directly in \acp{csd}. The extension corresponds to the residence time and the amplitude to the observed offset visible in \cref{fig:noise_examples}g.
    
    \subsubsection{White Noise}
    \label{sssec:white_parameters}
    In the \ac{psd} of a \ac{csd}, white noise dominates in the highest frequencies, while other noise types and the signal itself prevail in lower frequencies. The other types of noise and the signal itself dominate in lower frequencies. The \ac{psd} of white noise is given by
    \begin{equation}
    	\label{equ:psd_white}
    	\begin{gathered}
    	PSD_{w} = c_{w} \cdot \sigma_{w}^2.
    	\end{gathered}
    \end{equation}
    However, as the ratio between pink and white noise varies, we cannot determine a fixed corner frequency for the noise separation. Thus, we fit the sum
    \begin{equation}
    	PSD_{w,p} = c_{w} \cdot \sigma_{w}^2 + \frac{c_{p}}{f} \cdot \sigma_{p}^2
    \end{equation}
    of pink and white noise to the highest frequencies of the \ac{psd} computed by Welch's method \cite{welch} \footnote{$c_{p} = 0.1$ and $c_{w} = 2$ for the Welch method} from \texttt{SciPy} \cite{scipy}. The fit is applied to the average \ac{psd} of all rows to provide a better approximation.\\\\
    We tested our method with simulated data to study its accuracy. It becomes apparent that it works well for $\sigma_{w} > 0.001$ and tends to underestimate the intensity of white noise otherwise. 
    
    \section{Evaluation}
    \label{sec:evaluation}
    
    For assessment, we first visually evaluate our implemented simulation\footnote{We used the configuration \texttt{default\_{}configs["GaAs\_{}v1"]} provided in \cite{simcats_github}, but additionally with varying sensors from \cite{fleitmann_2023}.} concerning the fidelity, diversity, and plausibility of the generated \acp{csd}. Then, we measure the equivalency to experimental data and, finally, show the performance of our model compared to simulations based on physical models.
    
    \subsection{Visual assessment}
\begin{figure}[ht]
        \centering
        \includegraphics[]{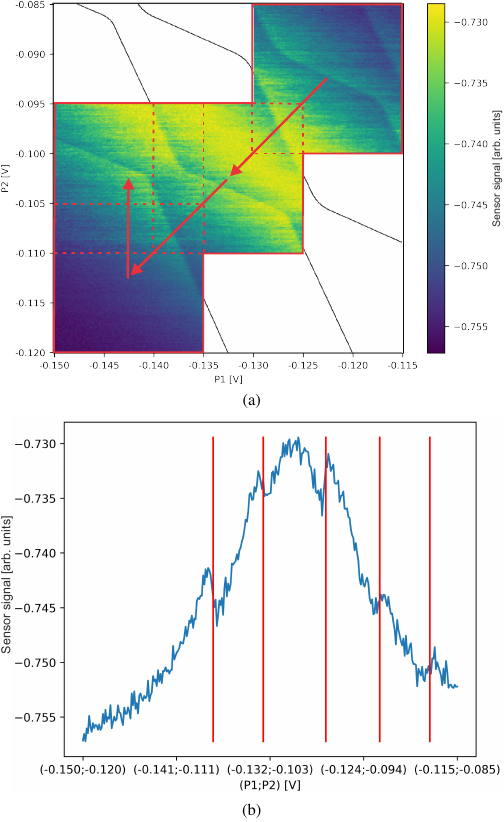}
        \caption{Example of simulation results. (a) Simulation of an iterative \ac{csd} measurement sequence during tuning. Red boxes illustrate the individually measured \acp{csd}, with the dotted part indicating their overlap and the red arrows their chronological order. (b) A diagonal 1D scan in the voltage space of (a) with red lines indicating the \acp{ldt}.}
        \label{fig:simulation_example}
    \end{figure}

    \cref{fig:simulation_example}a shows an example of a series of 2D simulations. The structures of the characteristic honeycombs change noticeably over the voltage range, matching the observations from experiments. Also, the distortions and the sensor response resemble the measurements.

    \subsection{Metrics for Generative Models}
    \label{ssec:metrics}
    The problem of comparing distributions of generated and measured data also appears in the context of machine learning when evaluating generative models. However, many of the used metrics in that field, e.g., Inception Score \cite{inception_score} and Fréchet Inception Distance \cite{frechet_distance}, are not applicable here because their classification method has to be trained application-specific or is pre-trained on natural images. Moreover, a metric computed sample-wise allows for a better analysis of the simulation model deficiencies. Therefore, we use the $\alpha$-precision and $\beta$-recall metrics to measure the fidelity and diversity of the generated datasets \cite{precision_recall}. \(\alpha\)-precision describes the probability that a generated sample exists in the $\alpha$-support of the measured data, whereas \(\beta\)-recall describes the fraction of measured samples that reside in the $\beta$-support of the generated data in combination with a chosen \(k\)-neighborhood\footnote{$\alpha$- or $\beta$-support is the minimum volume (sphere) subset of the whole set that supports a probability mass of $\alpha$ or $\beta$.}. Both metrics range from zero to one, with high values indicating similar distributions of measured and generated data. \cite{precision_recall} introduced a third metric that indicates if a generative model tends only to copy the training data. This metric is not crucial for our evaluation as we use only the determined parameter ranges for the generation. \\
    To compute the metrics, we first embed the \acp{csd} $x_i$ into a feature hypersphere. Therefore, we use an own adaptation of the \texttt{MNIST\_LeNet} neural network $\phi$ implemented in \cite{deepSVDD_implementation}. The training minimizes the loss \cite{deepSVDD_paper}
    \begin{equation}
    	L = R^2 + \frac{1}{\nu n} \sum_{i=1}^n \max\{ 0, \lVert \phi(x_i) - c\rVert^2 - R^2\},
    \end{equation}
    where $R$ is the radius of the hypersphere, $c$ represents the center, $\nu$ denotes a balancing factor, and $n$ is the number of data points. A corresponding framework for training neural networks is available on GitHub \cite{deepSVDD_implementation}.\\
    The selected hyperparameters for the training of the neural network and the computation of the metrics reside in \cite{fleitmann_2023}.
    
    \subsubsection{Preprocessing and Training}
    Before applying the neural network, a normalization aligns the different value ranges of the \acp{csd}. \\
    We train the network with randomly selected 50\%{} of the available 484 experimental data, with the other 50\%{} constituting the test set. In order to increase the amount of data during training, we apply rotations, flips, and random brightness and contrast changes to the \acp{csd} using the Python package \texttt{albumentations}\cite{albumentations}.
    
    \subsubsection{Results}
    For the result investigation, our simulation should include only all outliers represented in the experimental data. Thus, we set the parameters \(\alpha\) and \(\beta\) to 1. Then, we test whether our network can embed \acp{csd} into a feature hypersphere that sufficiently distinguishes \acp{csd} from non-\ac{csd} data. Therefore, we benchmark the training set against the retained experimental test set. It achieves high precision and recall values, indicating a reasonable mapping. For the \(k\)-nearest-neighbor region, we find \(k=9\) as the minimum, leading to a recall of 100\%{} in our data \cite{precision_recall}.\\
    Furthermore, we test whether the network embeds non-\ac{csd} data into the same hypersphere. Tests with MNIST data\footnote{The MNIST dataset consists of images containing handwritten digits.} \cite{mnist} achieve a high precision but a very low recall, indicating that they occupy a non-coinciding subspace. In conclusion, our network can map \ac{csd} data into an appropriate feature hypersphere, and the applied metrics are suitable to evaluate the equivalency of experimental and simulated \acp{csd}.\\
    The metrics result in a high precision and a recall of 67.5\%{}. In contrast to the simulated data, we must consider that the experimental data do not cover the whole voltage space homogeneously but prefer particular regions due to the experimenter's experience. A comparison with about ten times the amount of data increases the recall to 79.0\%{}, which supports our hypothesis. In contrast, using ten times more MNIST data than experimental \acp{csd} does not significantly increase the recall. \\
    In summary, the simulated data align strongly with the experimental data and map their distribution to a large extent. Nevertheless, the results indicate that our simulation does not yet represent all experimental \acp{csd}. As some of the available experimental datasets include anomalies from postprocessing steps not represented in our simulation, we expect an even higher coverage for unprocessed data.

    \begin{table}[t]
      \centering
      \caption{Evaluation results utilizing 1-precision and 1-recall. We compare the listed datasets to the experimental data test set. The datasets MNIST and SimCATS contain the same number of images as the experimental test set. Additionally, we supply results for expanded MNIST and SimCATS datasets with roughly ten times the size of the test set.}
      \label{table:metrics_results}
      \begin{tabularx}{\linewidth}{Xrr}
        \toprule
        \textbf{Dataset} & \textbf{1-precision [\%{}]} & \textbf{1-recall (k=9) [\%{}]} \\
        \midrule
        Experimental training set & 98.8 & 100.0 \\
        \midrule
        MNIST & 100.0 & 2.5 \\
        \midrule
        MNIST expanded & 100.0 & 3.3 \\
        \midrule
        SimCATS & 99.6 & 67.5 \\
        \midrule
        SimCATS expanded & 99.8 & 79.0 \\
        \bottomrule
      \end{tabularx}
    \end{table}

    \subsection{Performance analysis}
\begin{figure}[ht]
        \centering
        \includegraphics[width=\linewidth]{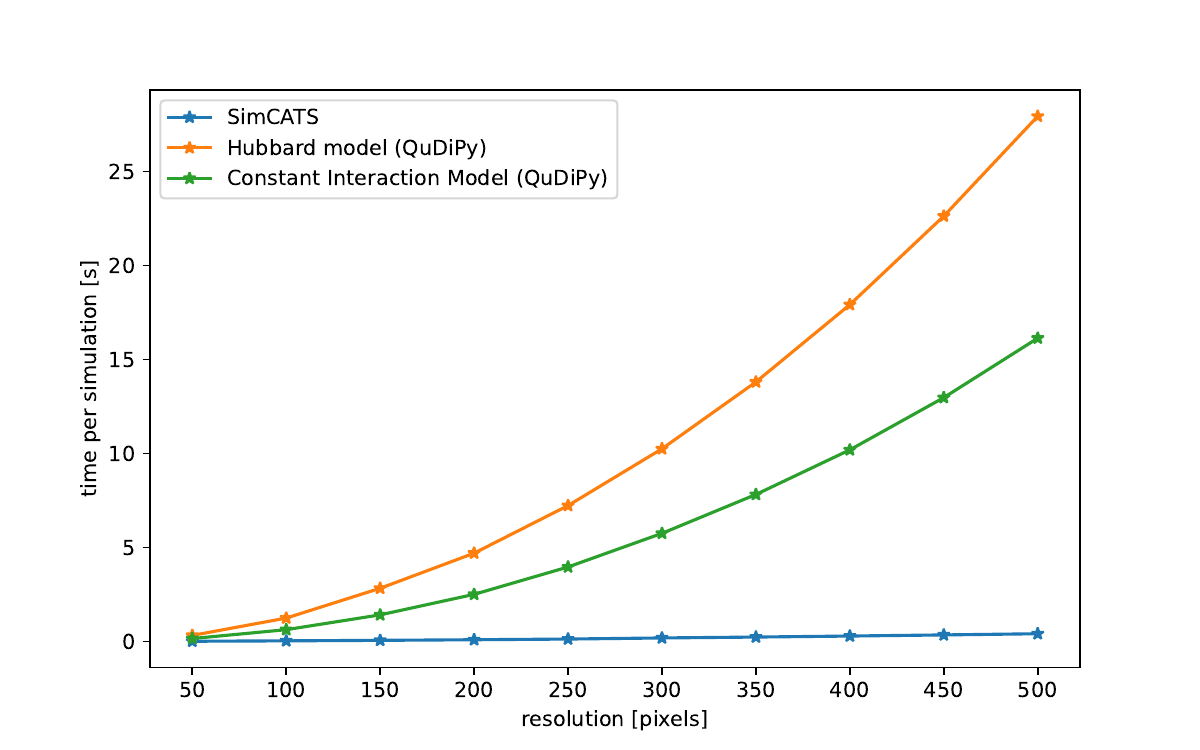}
        \caption{Average execution time of different simulation approaches in dependence on the resolution (in pixels per axis). The time refers to the computation per \ac{csd} (without distortions) after the previous initialization. Physical simulation models implemented in QuDiPy \cite{qudipy} show a quadratic time performance, while SimCATS hardly depends on the image resolution.}
        \label{fig:time_measurements}
    \end{figure}

    We benchmark the geometric simulation approach against two typical physical simulations regarding the execution time per simulation using an Intel Xeon w5-2455X - 3.19 GHz. For the comparison, we configure the parameters of the different approaches so that the simulated area covers similar structures. Then, we perform simulations of different resolutions and average the execution time over 50 runs each. \cref{fig:time_measurements} visualizes the results. It is noticeable that the execution time of the physical models increases quadratically with the resolution per axis. Furthermore, the Hubbard model, which includes quantum effects like tunnel coupling, is significantly slower than the constant interaction model. In comparison, the execution time for the geometric model is always lower\footnote{For the resolutions in \([50, 500]\), the execution time is in \([19, 418]\) \si{\milli\second} with a memory requirement in \([44.4, 90.0]\) MB.} and hardly increases because the computation of \acp{tct} is the decisive factor here, whereas the calculation of pixels is very fast. 

    \section{Conclusion}
    \label{sec:conclusion}
    We presented an approach for the simulation of \ac{csd} data, which incorporates the most relevant effects observed in measurements. First, we defined our model, comprising the ideal data generation, the sensor reaction, and distortions. Our \ac{tct}-based geometric representation of \ac{csd} structures enables the ideal data simulation for \acp{dqd}, independent of the sample material and layout. Furthermore, the sensor model includes the representation of observed Coulomb peaks as simplified Lorentzians and the sensor potential's dependency on lever arms of the \ac{dqd}'s gate voltages and occupation. Considering a typical measurement signal path, we assigned the distortions to three newly proposed categories and defined their source, simulation, and parameters afterward. Next, we extracted the different model parameters from measured data and, finally, showed simulation results and analyzed the capabilities of the simulation to mimic a diverse set of measured data.

    We suggest using our simulation framework for tuning algorithm development and benchmarking. It generates realistic data, allows a quick generation of large datasets with known ground truth, and enables fair comparability of diverse approaches of different sites. With complete control over the strength of the distortions and the sensitivity of the sensor, it additionally enables the determination of minimum measurement requirements for the success of an algorithm. Furthermore, using the provided interfaces, our open-source Python framework is designed for simple extension and adaptation. Thus, we highly encourage the contributions of other groups to build up a standard framework that drives the development of automated tuning solutions.
    
    Future work on our \ac{csd} simulation could incorporate further effects and their influence on the data. For example, computing wider \ac{csd} scans\footnote{For applications that analyze only small voltage ranges, we do not consider this to be necessary} can incorporate the varying lengths of the \acp{ldt}. 
    The correlation between consecutively recorded \acp{csd} or multiple \acp{dqd} measured at the same time are of minor impact but can be included if required.
    Another open question addresses the relationship between the geometric \ac{tct} parameters and the parameters of physical models. 
    With such knowledge, the simulation can directly adapt the geometric parameters to changes in the system. Thus, our fast simulation approach can support the development of more complex tuning routines. Furthermore, algorithms only trained on simulated data must be tested in experiments, and parameter sets by other sites’ experiments are necessary to develop and benchmark robust technology-independent algorithms.

    \section*{Acknowledgment}
    
    We appreciate helpful discussions with Mr. Jan-Oliver Kropp on evaluation metrics for generative models.

    \ifCLASSOPTIONcaptionsoff
      \newpage
    \fi
    
    \bibliographystyle{IEEEtran}
    \bibliography{IEEEabrv, References}

\begin{thebibliography}{100}
\providecommand{\url}[1]{#1}
\csname url@samestyle\endcsname
\providecommand{\newblock}{\relax}
\providecommand{\bibinfo}[2]{#2}
\providecommand{\BIBentrySTDinterwordspacing}{\spaceskip=0pt\relax}
\providecommand{\BIBentryALTinterwordstretchfactor}{4}
\providecommand{\BIBentryALTinterwordspacing}{\spaceskip=\fontdimen2\font plus
\BIBentryALTinterwordstretchfactor\fontdimen3\font minus
  \fontdimen4\font\relax}
\providecommand{\BIBforeignlanguage}[2]{{%
\expandafter\ifx\csname l@#1\endcsname\relax
\typeout{** WARNING: IEEEtran.bst: No hyphenation pattern has been}%
\typeout{** loaded for the language `#1'. Using the pattern for}%
\typeout{** the default language instead.}%
\else
\language=\csname l@#1\endcsname
\fi
#2}}
\providecommand{\BIBdecl}{\relax}
\BIBdecl

\bibitem{grabert_1992}
\BIBentryALTinterwordspacing
H.~Grabert and M.~H. Devoret, Eds., \emph{Single {{Charge Tunneling}}}, ser.
  {{NATO ASI Series}}.\hskip 1em plus 0.5em minus 0.4em\relax {Boston, MA}:
  {Springer US}, 1992, vol. 294. [Online]. Available:
  \url{http://link.springer.com/10.1007/978-1-4757-2166-9}
\BIBentrySTDinterwordspacing

\bibitem{nazarov_1993}
\BIBentryALTinterwordspacing
Y.~V. Nazarov, ``Quantum interference, tunnel junctions and resonant tunneling
  interferometer,'' \emph{Physica B: Condensed Matter}, vol. 189, no.~1, pp.
  57--69, Jun. 1993, \href{https://doi.org/10.1016/0921-4526(93)90146-W}{doi:
  10.1016/0921-4526(93)90146-W}. [Online]. Available:
  \url{https://www.sciencedirect.com/science/article/pii/092145269390146W}
\BIBentrySTDinterwordspacing

\bibitem{konig_1996}
\BIBentryALTinterwordspacing
J.~K{\"o}nig, J.~Schmid, H.~Schoeller, and G.~Sch{\"o}n, ``Resonant tunneling
  through ultrasmall quantum dots: {{Zero-bias}} anomalies, magnetic-field
  dependence, and boson-assisted transport,'' \emph{Physical Review B},
  vol.~54, no.~23, pp. 16\,820--16\,837, Dec. 1996,
  \href{https://doi.org/10.1103/PhysRevB.54.16820}{doi:
  10.1103/PhysRevB.54.16820}. [Online]. Available:
  \url{https://link.aps.org/doi/10.1103/PhysRevB.54.16820}
\BIBentrySTDinterwordspacing

\bibitem{timm_2008}
\BIBentryALTinterwordspacing
C.~Timm, ``Tunneling through molecules and quantum dots: {{Master-equation}}
  approaches,'' \emph{Physical Review B}, vol.~77, no.~19, p. 195416, May 2008,
  \href{https://doi.org/10.1103/PhysRevB.77.195416}{doi:
  10.1103/PhysRevB.77.195416}. [Online]. Available:
  \url{https://link.aps.org/doi/10.1103/PhysRevB.77.195416}
\BIBentrySTDinterwordspacing

\bibitem{feynman_1963}
\BIBentryALTinterwordspacing
R.~P. Feynman and F.~L. Vernon, ``The theory of a general quantum system
  interacting with a linear dissipative system,'' \emph{Annals of Physics},
  vol.~24, pp. 118--173, Oct. 1963,
  \href{https://doi.org/10.1016/0003-4916(63)90068-X}{doi:
  10.1016/0003-4916(63)90068-X}. [Online]. Available:
  \url{https://www.sciencedirect.com/science/article/pii/000349166390068X}
\BIBentrySTDinterwordspacing

\bibitem{singh_1994}
\BIBentryALTinterwordspacing
D.~J. Singh, ``Overview of {{Density Functional Theory}} and {{Methods}},'' in
  \emph{Planewaves, {{Pseudopotentials}} and the {{LAPW Method}}}, D.~J. Singh,
  Ed.\hskip 1em plus 0.5em minus 0.4em\relax {Boston, MA}: {Springer US}, 1994,
  pp. 5--16. [Online]. Available:
  \url{https://doi.org/10.1007/978-1-4757-2312-0_2}
\BIBentrySTDinterwordspacing

\bibitem{siano_2004}
\BIBentryALTinterwordspacing
F.~Siano and R.~Egger, ``Josephson {{Current}} through a {{Nanoscale Magnetic
  Quantum Dot}},'' \emph{Physical Review Letters}, vol.~93, no.~4, p. 047002,
  Jul. 2004, \href{https://doi.org/10.1103/PhysRevLett.93.047002}{doi:
  10.1103/PhysRevLett.93.047002}. [Online]. Available:
  \url{https://link.aps.org/doi/10.1103/PhysRevLett.93.047002}
\BIBentrySTDinterwordspacing

\bibitem{pokorny_2018}
\BIBentryALTinterwordspacing
V.~Pokorn{\'y} and M.~{\v Z}onda, ``Correlation effects in superconducting
  quantum dot systems,'' \emph{Physica B: Condensed Matter}, vol. 536, pp.
  488--491, May 2018, \href{https://doi.org/10.1016/j.physb.2017.08.059}{doi:
  10.1016/j.physb.2017.08.059}. [Online]. Available:
  \url{https://www.sciencedirect.com/science/article/pii/S0921452617305446}
\BIBentrySTDinterwordspacing

\bibitem{hepp_1974}
\BIBentryALTinterwordspacing
K.~Hepp, ``The classical limit for quantum mechanical correlation functions,''
  \emph{Communications in Mathematical Physics}, vol.~35, no.~4, pp. 265--277,
  Dec. 1974, \href{https://doi.org/10.1007/BF01646348}{doi:
  10.1007/BF01646348}. [Online]. Available:
  \url{https://doi.org/10.1007/BF01646348}
\BIBentrySTDinterwordspacing

\bibitem{ginibre_1979}
\BIBentryALTinterwordspacing
J.~Ginibre and G.~Velo, ``The classical field limit of scattering theory for
  non-relativistic many-boson systems. {{II}},'' \emph{Communications in
  Mathematical Physics}, vol.~68, no.~1, pp. 45--68, Feb. 1979,
  \href{https://doi.org/10.1007/BF01562541}{doi: 10.1007/BF01562541}. [Online].
  Available: \url{https://doi.org/10.1007/BF01562541}
\BIBentrySTDinterwordspacing

\bibitem{erdos_2001}
\BIBentryALTinterwordspacing
L.~Erd{\H o}s and H.-T. Yau, ``Derivation of the nonlinear {{Schr\"odinger}}
  equation from a many-body {{Coulomb}} system,'' \emph{Advances in Theoretical
  and Mathematical Physics}, vol.~5, no.~6, pp. 1169--1205, 2001,
  \href{https://doi.org/10.4310/ATMP.2001.v5.n6.a6}{doi:
  10.4310/ATMP.2001.v5.n6.a6}. [Online]. Available:
  \url{http://www.intlpress.com/site/pub/pages/journals/items/atmp/content/vols/0005/0006/a006/}
\BIBentrySTDinterwordspacing

\bibitem{bulka_2004}
\BIBentryALTinterwordspacing
B.~R. Bu{\l}ka and T.~Kostyrko, ``Electronic correlations in coherent transport
  through a two quantum dot system,'' \emph{Physical Review B}, vol.~70,
  no.~20, p. 205333, Nov. 2004,
  \href{https://doi.org/10.1103/PhysRevB.70.205333}{doi:
  10.1103/PhysRevB.70.205333}. [Online]. Available:
  \url{https://link.aps.org/doi/10.1103/PhysRevB.70.205333}
\BIBentrySTDinterwordspacing

\bibitem{sztenkiel_2007}
\BIBentryALTinterwordspacing
D.~Sztenkiel and R.~{\'S}wirkowicz, ``Electron {{Transport}} through {{Double
  Quantum Dot System}} with {{Inter-Dot Coulomb Interaction}},'' \emph{Acta
  Physica Polonica A}, vol. 111, no.~3, pp. 361--372, Mar. 2007,
  \href{https://doi.org/10.12693/APhysPolA.111.361}{doi:
  10.12693/APhysPolA.111.361}. [Online]. Available:
  \url{http://przyrbwn.icm.edu.pl/APP/PDF/111/a111z306.pdf}
\BIBentrySTDinterwordspacing

\bibitem{zagoskin_2014}
\BIBentryALTinterwordspacing
A.~Zagoskin, \emph{Quantum {{Theory}} of {{Many-Body Systems}}: {{Techniques}}
  and {{Applications}}}, ser. Graduate {{Texts}} in {{Physics}}.\hskip 1em plus
  0.5em minus 0.4em\relax {Cham}: {Springer International Publishing}, 2014.
  [Online]. Available:
  \url{https://link.springer.com/10.1007/978-3-319-07049-0}
\BIBentrySTDinterwordspacing

\bibitem{brody_1981}
\BIBentryALTinterwordspacing
T.~A. Brody, J.~Flores, J.~B. French, P.~A. Mello, A.~Pandey, and S.~S.~M.
  Wong, ``Random-matrix physics: Spectrum and strength fluctuations,''
  \emph{Reviews of Modern Physics}, vol.~53, no.~3, pp. 385--479, Jul. 1981,
  \href{https://doi.org/10.1103/RevModPhys.53.385}{doi:
  10.1103/RevModPhys.53.385}. [Online]. Available:
  \url{https://link.aps.org/doi/10.1103/RevModPhys.53.385}
\BIBentrySTDinterwordspacing

\bibitem{newton_1982}
\BIBentryALTinterwordspacing
R.~G. Newton, \emph{Scattering {{Theory}} of {{Waves}} and
  {{Particles}}}.\hskip 1em plus 0.5em minus 0.4em\relax {Berlin, Heidelberg}:
  {Springer}, 1982. [Online]. Available:
  \url{http://link.springer.com/10.1007/978-3-642-88128-2}
\BIBentrySTDinterwordspacing

\bibitem{beenakker_1997}
\BIBentryALTinterwordspacing
C.~W.~J. Beenakker, ``Random-matrix theory of quantum transport,''
  \emph{Reviews of Modern Physics}, vol.~69, no.~3, pp. 731--808, Jul. 1997,
  \href{https://doi.org/10.1103/RevModPhys.69.731}{doi:
  10.1103/RevModPhys.69.731}. [Online]. Available:
  \url{https://link.aps.org/doi/10.1103/RevModPhys.69.731}
\BIBentrySTDinterwordspacing

\bibitem{potz_2008}
\BIBentryALTinterwordspacing
W.~P{\"o}tz, ``Scattering approach to semiconductor double quantum dot
  cavities,'' \emph{Physical Review B}, vol.~77, no.~3, p. 035310, Jan. 2008,
  \href{https://doi.org/10.1103/PhysRevB.77.035310}{doi:
  10.1103/PhysRevB.77.035310}. [Online]. Available:
  \url{https://link.aps.org/doi/10.1103/PhysRevB.77.035310}
\BIBentrySTDinterwordspacing

\bibitem{feynman_1982}
\BIBentryALTinterwordspacing
R.~P. Feynman, ``Simulating physics with computers,'' \emph{International
  Journal of Theoretical Physics}, vol.~21, no.~6, pp. 467--488, Jun. 1982,
  \href{https://doi.org/10.1007/BF02650179}{doi: 10.1007/BF02650179}. [Online].
  Available: \url{https://doi.org/10.1007/BF02650179}
\BIBentrySTDinterwordspacing

\bibitem{georgescu_2014}
\BIBentryALTinterwordspacing
I.~M. Georgescu, S.~Ashhab, and F.~Nori, ``Quantum simulation,'' \emph{Reviews
  of Modern Physics}, vol.~86, no.~1, pp. 153--185, Mar. 2014,
  \href{https://doi.org/10.1103/RevModPhys.86.153}{doi:
  10.1103/RevModPhys.86.153}. [Online]. Available:
  \url{https://link.aps.org/doi/10.1103/RevModPhys.86.153}
\BIBentrySTDinterwordspacing

\bibitem{vanderwiel_2002}
\BIBentryALTinterwordspacing
W.~G. {van der Wiel}, S.~De~Franceschi, J.~M. Elzerman, T.~Fujisawa,
  S.~Tarucha, and L.~P. Kouwenhoven, ``Electron transport through double
  quantum dots,'' \emph{Reviews of Modern Physics}, vol.~75, no.~1, pp. 1--22,
  Dec. 2002, \href{https://doi.org/10.1103/RevModPhys.75.1}{doi:
  10.1103/RevModPhys.75.1}. [Online]. Available:
  \url{http://arxiv.org/abs/cond-mat/0205350}
\BIBentrySTDinterwordspacing

\bibitem{hanson_2007}
\BIBentryALTinterwordspacing
R.~Hanson, L.~P. Kouwenhoven, J.~R. Petta, S.~Tarucha, and L.~M.~K.
  Vandersypen, ``Spins in few-electron quantum dots,'' \emph{Reviews of Modern
  Physics}, vol.~79, no.~4, pp. 1217--1265, Oct. 2007,
  \href{https://doi.org/10.1103/RevModPhys.79.1217}{doi:
  10.1103/RevModPhys.79.1217}. [Online]. Available:
  \url{http://arxiv.org/abs/cond-mat/0610433}
\BIBentrySTDinterwordspacing

\bibitem{barthelemy_2013}
\BIBentryALTinterwordspacing
P.~Barthelemy and L.~M.~K. Vandersypen, ``Quantum {{Dot Systems}}: A versatile
  platform for quantum simulations,'' \emph{Annalen der Physik}, vol. 525, no.
  10-11, pp. 808--826, Nov. 2013,
  \href{https://doi.org/10.1002/andp.201300124}{doi: 10.1002/andp.201300124}.
  [Online]. Available:
  \url{https://onlinelibrary.wiley.com/doi/10.1002/andp.201300124}
\BIBentrySTDinterwordspacing

\bibitem{byrnes_2008}
\BIBentryALTinterwordspacing
T.~Byrnes, N.~Y. Kim, K.~Kusudo, and Y.~Yamamoto, ``Quantum simulation of
  {{Fermi-Hubbard}} models in semiconductor quantum-dot arrays,''
  \emph{Physical Review B}, vol.~78, no.~7, p. 075320, Aug. 2008,
  \href{https://doi.org/10.1103/PhysRevB.78.075320}{doi:
  10.1103/PhysRevB.78.075320}. [Online]. Available:
  \url{https://link.aps.org/doi/10.1103/PhysRevB.78.075320}
\BIBentrySTDinterwordspacing

\bibitem{hensgens_2017}
\BIBentryALTinterwordspacing
T.~Hensgens, T.~Fujita, L.~Janssen, X.~Li, C.~J. Van~Diepen, C.~Reichl,
  W.~Wegscheider, S.~D. Sarma, and L.~M.~K. Vandersypen, ``Quantum simulation
  of a {{Fermi-Hubbard}} model using a semiconductor quantum dot array,''
  \emph{Nature}, vol. 548, no. 7665, pp. 70--73, Aug. 2017,
  \href{https://doi.org/10.1038/nature23022}{doi: 10.1038/nature23022}.
  [Online]. Available: \url{http://arxiv.org/abs/1702.07511}
\BIBentrySTDinterwordspacing

\bibitem{hensgens_2018}
\BIBentryALTinterwordspacing
T.~Hensgens, ``Emulating {{Fermi-Hubbard}} physics with quantum dots,'' Ph.D.
  dissertation, Delft University of Technology, 2018. [Online]. Available:
  \url{http://resolver.tudelft.nl/uuid:b71f3b0b-73a0-4996-896c-84ed43e72035}
\BIBentrySTDinterwordspacing

\bibitem{wang_2022}
\BIBentryALTinterwordspacing
X.~Wang, E.~Khatami, F.~Fei, J.~Wyrick, P.~Namboodiri, R.~Kashid, A.~F. Rigosi,
  G.~Bryant, and R.~Silver, ``Experimental realization of an extended
  {{Fermi-Hubbard}} model using a {{2D}} lattice of dopant-based quantum
  dots,'' \emph{Nature Communications}, vol.~13, no.~1, p. 6824, Nov. 2022,
  \href{https://doi.org/10.1038/s41467-022-34220-w}{doi:
  10.1038/s41467-022-34220-w}. [Online]. Available:
  \url{https://www.nature.com/articles/s41467-022-34220-w}
\BIBentrySTDinterwordspacing

\bibitem{dehollain_2020}
\BIBentryALTinterwordspacing
J.~P. Dehollain, U.~Mukhopadhyay, V.~P. Michal, Y.~Wang, B.~Wunsch, C.~Reichl,
  W.~Wegscheider, M.~S. Rudner, E.~Demler, and L.~M.~K. Vandersypen, ``Nagaoka
  ferromagnetism observed in a quantum dot plaquette,'' \emph{Nature}, vol.
  579, no. 7800, pp. 528--533, Mar. 2020,
  \href{https://doi.org/10.1038/s41586-020-2051-0}{doi:
  10.1038/s41586-020-2051-0}. [Online]. Available:
  \url{https://www.nature.com/articles/s41586-020-2051-0}
\BIBentrySTDinterwordspacing

\bibitem{vandiepen_2021}
\BIBentryALTinterwordspacing
C.~{van Diepen,}, ``Quantum simulation with electron spins in quantum dots,''
  Ph.D. dissertation, Delft University of Technology, 2021. [Online].
  Available:
  \url{http://resolver.tudelft.nl/uuid:f7adb947-1d56-4326-aa99-13b5baa353f6}
\BIBentrySTDinterwordspacing

\bibitem{vandiepen_2021a}
\BIBentryALTinterwordspacing
C.~J. {van Diepen}, T.-K. Hsiao, U.~Mukhopadhyay, C.~Reichl, W.~Wegscheider,
  and L.~M.~K. Vandersypen, ``Quantum {{Simulation}} of {{Antiferromagnetic
  Heisenberg Chain}} with {{Gate-Defined Quantum Dots}},'' \emph{Physical
  Review X}, vol.~11, no.~4, p. 041025, Nov. 2021,
  \href{https://doi.org/10.1103/PhysRevX.11.041025}{doi:
  10.1103/PhysRevX.11.041025}. [Online]. Available:
  \url{https://link.aps.org/doi/10.1103/PhysRevX.11.041025}
\BIBentrySTDinterwordspacing

\bibitem{suzuki_1993}
\BIBentryALTinterwordspacing
M.~Suzuki, \emph{Quantum {{Monte Carlo Methods}} in {{Condensed Matter
  Physics}}}.\hskip 1em plus 0.5em minus 0.4em\relax {WORLD SCIENTIFIC}, Dec.
  1993. [Online]. Available:
  \url{https://www.worldscientific.com/worldscibooks/10.1142/2262}
\BIBentrySTDinterwordspacing

\bibitem{luitz_2010}
\BIBentryALTinterwordspacing
D.~J. Luitz and F.~F. Assaad, ``Weak-coupling continuous-time quantum {{Monte
  Carlo}} study of the single impurity and periodic {{Anderson}} models with
  \$s\$-wave superconducting baths,'' \emph{Physical Review B}, vol.~81, no.~2,
  p. 024509, Jan. 2010, \href{https://doi.org/10.1103/PhysRevB.81.024509}{doi:
  10.1103/PhysRevB.81.024509}. [Online]. Available:
  \url{https://link.aps.org/doi/10.1103/PhysRevB.81.024509}
\BIBentrySTDinterwordspacing

\bibitem{dreizler_1990}
\BIBentryALTinterwordspacing
R.~M. Dreizler and E.~K.~U. Gross, ``Many-{{Body Perturbation Theory}},'' in
  \emph{Density {{Functional Theory}}: {{An Approach}} to the {{Quantum
  Many-Body Problem}}}, R.~M. Dreizler and E.~K.~U. Gross, Eds.\hskip 1em plus
  0.5em minus 0.4em\relax {Berlin, Heidelberg}: {Springer}, 1990, pp. 138--172.
  [Online]. Available: \url{https://doi.org/10.1007/978-3-642-86105-5_6}
\BIBentrySTDinterwordspacing

\bibitem{fetter_2003}
A.~L. Fetter and J.~D. Walecka, \emph{Quantum {{Theory}} of {{Many-particle
  Systems}}}.\hskip 1em plus 0.5em minus 0.4em\relax {Courier Corporation},
  Jun. 2003.

\bibitem{meyer_1990}
\BIBentryALTinterwordspacing
H.~D. Meyer, U.~Manthe, and L.~S. Cederbaum, ``The multi-configurational
  time-dependent {{Hartree}} approach,'' \emph{Chemical Physics Letters}, vol.
  165, no.~1, pp. 73--78, Jan. 1990,
  \href{https://doi.org/10.1016/0009-2614(90)87014-I}{doi:
  10.1016/0009-2614(90)87014-I}. [Online]. Available:
  \url{https://www.sciencedirect.com/science/article/pii/000926149087014I}
\BIBentrySTDinterwordspacing

\bibitem{manthe_1992}
\BIBentryALTinterwordspacing
U.~Manthe, H.-D. Meyer, and L.~S. Cederbaum, ``Wave-packet dynamics within the
  multiconfiguration {{Hartree}} framework: {{General}} aspects and application
  to {{NOCl}},'' \emph{The Journal of Chemical Physics}, vol.~97, no.~5, pp.
  3199--3213, Sep. 1992, \href{https://doi.org/10.1063/1.463007}{doi:
  10.1063/1.463007}. [Online]. Available:
  \url{https://doi.org/10.1063/1.463007}
\BIBentrySTDinterwordspacing

\bibitem{beck_2000}
\BIBentryALTinterwordspacing
M.~H. Beck, A.~J{\"a}ckle, G.~A. Worth, and H.~D. Meyer, ``The
  multiconfiguration time-dependent {{Hartree}} ({{MCTDH}}) method: A highly
  efficient algorithm for propagating wavepackets,'' \emph{Physics Reports},
  vol. 324, no.~1, pp. 1--105, Jan. 2000,
  \href{https://doi.org/10.1016/S0370-1573(99)00047-2}{doi:
  10.1016/S0370-1573(99)00047-2}. [Online]. Available:
  \url{https://www.sciencedirect.com/science/article/pii/S0370157399000472}
\BIBentrySTDinterwordspacing

\bibitem{tanimura_1989}
\BIBentryALTinterwordspacing
Y.~Tanimura and R.~Kubo, ``Time {{Evolution}} of a {{Quantum System}} in
  {{Contact}} with a {{Nearly Gaussian-Markoffian Noise Bath}},'' \emph{Journal
  of the Physical Society of Japan}, vol.~58, no.~1, pp. 101--114, Jan. 1989,
  \href{https://doi.org/10.1143/JPSJ.58.101}{doi: 10.1143/JPSJ.58.101}.
  [Online]. Available: \url{https://journals.jps.jp/doi/10.1143/JPSJ.58.101}
\BIBentrySTDinterwordspacing

\bibitem{tanimura_1990}
\BIBentryALTinterwordspacing
Y.~Tanimura, ``Nonperturbative expansion method for a quantum system coupled to
  a harmonic-oscillator bath,'' \emph{Physical Review A}, vol.~41, no.~12, pp.
  6676--6687, Jun. 1990, \href{https://doi.org/10.1103/PhysRevA.41.6676}{doi:
  10.1103/PhysRevA.41.6676}. [Online]. Available:
  \url{https://link.aps.org/doi/10.1103/PhysRevA.41.6676}
\BIBentrySTDinterwordspacing

\bibitem{tanimura_2006}
\BIBentryALTinterwordspacing
------, ``Stochastic {{Liouville}}, {{Langevin}},
  {{Fokker}}\textendash{{Planck}}, and {{Master Equation Approaches}} to
  {{Quantum Dissipative Systems}},'' \emph{Journal of the Physical Society of
  Japan}, vol.~75, no.~8, p. 082001, Aug. 2006,
  \href{https://doi.org/10.1143/JPSJ.75.082001}{doi: 10.1143/JPSJ.75.082001}.
  [Online]. Available: \url{https://journals.jps.jp/doi/10.1143/JPSJ.75.082001}
\BIBentrySTDinterwordspacing

\bibitem{carleo_2017}
\BIBentryALTinterwordspacing
G.~Carleo and M.~Troyer, ``Solving the quantum many-body problem with
  artificial neural networks,'' \emph{Science}, vol. 355, no. 6325, pp.
  602--606, Feb. 2017, \href{https://doi.org/10.1126/science.aag2302}{doi:
  10.1126/science.aag2302}. [Online]. Available:
  \url{https://www.science.org/doi/10.1126/science.aag2302}
\BIBentrySTDinterwordspacing

\bibitem{hartmann_2019}
\BIBentryALTinterwordspacing
M.~J. Hartmann and G.~Carleo, ``Neural-{{Network Approach}} to {{Dissipative
  Quantum Many-Body Dynamics}},'' \emph{Physical Review Letters}, vol. 122,
  no.~25, p. 250502, Jun. 2019,
  \href{https://doi.org/10.1103/PhysRevLett.122.250502}{doi:
  10.1103/PhysRevLett.122.250502}. [Online]. Available:
  \url{https://link.aps.org/doi/10.1103/PhysRevLett.122.250502}
\BIBentrySTDinterwordspacing

\bibitem{schutt_2019}
\BIBentryALTinterwordspacing
K.~T. Sch{\"u}tt, M.~Gastegger, A.~Tkatchenko, K.-R. M{\"u}ller, and R.~J.
  Maurer, ``Unifying machine learning and quantum chemistry with a deep neural
  network for molecular wavefunctions,'' \emph{Nature Communications}, vol.~10,
  no.~1, p. 5024, Nov. 2019,
  \href{https://doi.org/10.1038/s41467-019-12875-2}{doi:
  10.1038/s41467-019-12875-2}. [Online]. Available:
  \url{https://www.nature.com/articles/s41467-019-12875-2}
\BIBentrySTDinterwordspacing

\bibitem{hermann_2020}
\BIBentryALTinterwordspacing
J.~Hermann, Z.~Sch{\"a}tzle, and F.~No{\'e}, ``Deep-neural-network solution of
  the electronic {{Schr\"odinger}} equation,'' \emph{Nature Chemistry},
  vol.~12, no.~10, pp. 891--897, Oct. 2020,
  \href{https://doi.org/10.1038/s41557-020-0544-y}{doi:
  10.1038/s41557-020-0544-y}. [Online]. Available:
  \url{https://www.nature.com/articles/s41557-020-0544-y}
\BIBentrySTDinterwordspacing

\bibitem{manzhos_2020}
\BIBentryALTinterwordspacing
S.~Manzhos, ``Machine learning for the solution of the {{Schr\"odinger}}
  equation,'' \emph{Machine Learning: Science and Technology}, vol.~1, no.~1,
  p. 013002, Apr. 2020, \href{https://doi.org/10.1088/2632-2153/ab7d30}{doi:
  10.1088/2632-2153/ab7d30}. [Online]. Available:
  \url{https://dx.doi.org/10.1088/2632-2153/ab7d30}
\BIBentrySTDinterwordspacing

\bibitem{vidal_2004}
\BIBentryALTinterwordspacing
G.~Vidal, ``Efficient {{Simulation}} of {{One-Dimensional Quantum Many-Body
  Systems}},'' \emph{Physical Review Letters}, vol.~93, no.~4, p. 040502, Jul.
  2004, \href{https://doi.org/10.1103/PhysRevLett.93.040502}{doi:
  10.1103/PhysRevLett.93.040502}. [Online]. Available:
  \url{https://link.aps.org/doi/10.1103/PhysRevLett.93.040502}
\BIBentrySTDinterwordspacing

\bibitem{hong_2022}
\BIBentryALTinterwordspacing
R.~Hong, Y.-X. Xiao, J.~Hu, A.-C. Ji, and S.-J. Ran, ``Functional {{Tensor
  Network Solving Many-body Schr}}\textbackslash "odinger {{Equation}},''
  \emph{Physical Review B}, vol. 105, no.~16, p. 165116, Apr. 2022,
  \href{https://doi.org/10.1103/PhysRevB.105.165116}{doi:
  10.1103/PhysRevB.105.165116}. [Online]. Available:
  \url{http://arxiv.org/abs/2201.12823}
\BIBentrySTDinterwordspacing

\bibitem{orus_2019}
\BIBentryALTinterwordspacing
R.~Or{\'u}s, ``Tensor networks for complex quantum systems,'' \emph{Nature
  Reviews Physics}, vol.~1, no.~9, pp. 538--550, Sep. 2019,
  \href{https://doi.org/10.1038/s42254-019-0086-7}{doi:
  10.1038/s42254-019-0086-7}. [Online]. Available:
  \url{https://www.nature.com/articles/s42254-019-0086-7}
\BIBentrySTDinterwordspacing

\bibitem{wilson_1975}
\BIBentryALTinterwordspacing
K.~G. Wilson, ``The renormalization group: {{Critical}} phenomena and the
  {{Kondo}} problem,'' \emph{Reviews of Modern Physics}, vol.~47, no.~4, pp.
  773--840, Oct. 1975, \href{https://doi.org/10.1103/RevModPhys.47.773}{doi:
  10.1103/RevModPhys.47.773}. [Online]. Available:
  \url{https://link.aps.org/doi/10.1103/RevModPhys.47.773}
\BIBentrySTDinterwordspacing

\bibitem{yoshioka_2000}
\BIBentryALTinterwordspacing
T.~Yoshioka and Y.~Ohashi, ``Numerical {{Renormalization Group Studies}} on
  {{Single Impurity Anderson}} {{Model}} in {{Superconductivity}}: {{A Unified
  Treatment}} of {{Magnetic}}, {{Nonmagnetic Impurities}}, and {{Resonance
  Scattering}},'' \emph{Journal of the Physical Society of Japan}, vol.~69,
  no.~6, pp. 1812--1823, Jun. 2000,
  \href{https://doi.org/10.1143/JPSJ.69.1812}{doi: 10.1143/JPSJ.69.1812}.
  [Online]. Available: \url{https://journals.jps.jp/doi/10.1143/JPSJ.69.1812}
\BIBentrySTDinterwordspacing

\bibitem{oguri_2004}
\BIBentryALTinterwordspacing
A.~Oguri, Y.~Tanaka, and A.~C.~Hewson, ``Quantum {{Phase Transition}} in a
  {{Minimal Model}} for the {{Kondo Effect}} in a {{Josephson Junction}},''
  \emph{Journal of the Physical Society of Japan}, vol.~73, no.~9, pp.
  2494--2504, Sep. 2004, \href{https://doi.org/10.1143/JPSJ.73.2494}{doi:
  10.1143/JPSJ.73.2494}. [Online]. Available:
  \url{https://journals.jps.jp/doi/10.1143/JPSJ.73.2494}
\BIBentrySTDinterwordspacing

\bibitem{bulla_2008}
\BIBentryALTinterwordspacing
R.~Bulla, T.~A. Costi, and T.~Pruschke, ``Numerical renormalization group
  method for quantum impurity systems,'' \emph{Reviews of Modern Physics},
  vol.~80, no.~2, pp. 395--450, Apr. 2008,
  \href{https://doi.org/10.1103/RevModPhys.80.395}{doi:
  10.1103/RevModPhys.80.395}. [Online]. Available:
  \url{https://link.aps.org/doi/10.1103/RevModPhys.80.395}
\BIBentrySTDinterwordspacing

\bibitem{oguri_2013}
\BIBentryALTinterwordspacing
A.~Oguri, Y.~Tanaka, and J.~Bauer, ``Interplay between {{Kondo}} and
  {{Andreev-Josephson}} effects in a quantum dot coupled to one normal and two
  superconducting leads,'' \emph{Physical Review B}, vol.~87, no.~7, p. 075432,
  Feb. 2013, \href{https://doi.org/10.1103/PhysRevB.87.075432}{doi:
  10.1103/PhysRevB.87.075432}. [Online]. Available:
  \url{https://link.aps.org/doi/10.1103/PhysRevB.87.075432}
\BIBentrySTDinterwordspacing

\bibitem{fannes_1992}
\BIBentryALTinterwordspacing
M.~Fannes, B.~Nachtergaele, and R.~F. Werner, ``Finitely correlated states on
  quantum spin chains,'' \emph{Communications in Mathematical Physics}, vol.
  144, no.~3, pp. 443--490, Mar. 1992,
  \href{https://doi.org/10.1007/BF02099178}{doi: 10.1007/BF02099178}. [Online].
  Available: \url{https://doi.org/10.1007/BF02099178}
\BIBentrySTDinterwordspacing

\bibitem{vidal_2003}
\BIBentryALTinterwordspacing
G.~Vidal, ``Efficient {{Classical Simulation}} of {{Slightly Entangled Quantum
  Computations}},'' \emph{Physical Review Letters}, vol.~91, no.~14, p. 147902,
  Oct. 2003, \href{https://doi.org/10.1103/PhysRevLett.91.147902}{doi:
  10.1103/PhysRevLett.91.147902}. [Online]. Available:
  \url{https://link.aps.org/doi/10.1103/PhysRevLett.91.147902}
\BIBentrySTDinterwordspacing

\bibitem{verstraete_2008}
\BIBentryALTinterwordspacing
F.~Verstraete, V.~Murg, and J.~Cirac, ``Matrix product states, projected
  entangled pair states, and variational renormalization group methods for
  quantum spin systems,'' \emph{Advances in Physics}, vol.~57, no.~2, pp.
  143--224, Mar. 2008, \href{https://doi.org/10.1080/14789940801912366}{doi:
  10.1080/14789940801912366}. [Online]. Available:
  \url{https://doi.org/10.1080/14789940801912366}
\BIBentrySTDinterwordspacing

\bibitem{cirac_2021}
\BIBentryALTinterwordspacing
I.~Cirac, D.~{Perez-Garcia}, N.~Schuch, and F.~Verstraete, ``Matrix {{Product
  States}} and {{Projected Entangled Pair States}}: {{Concepts}},
  {{Symmetries}}, and {{Theorems}},'' \emph{Reviews of Modern Physics},
  vol.~93, no.~4, p. 045003, Dec. 2021,
  \href{https://doi.org/10.1103/RevModPhys.93.045003}{doi:
  10.1103/RevModPhys.93.045003}. [Online]. Available:
  \url{http://arxiv.org/abs/2011.12127}
\BIBentrySTDinterwordspacing

\bibitem{white_1992}
\BIBentryALTinterwordspacing
S.~R. White, ``Density matrix formulation for quantum renormalization groups,''
  \emph{Physical Review Letters}, vol.~69, no.~19, pp. 2863--2866, Nov. 1992,
  \href{https://doi.org/10.1103/PhysRevLett.69.2863}{doi:
  10.1103/PhysRevLett.69.2863}. [Online]. Available:
  \url{https://link.aps.org/doi/10.1103/PhysRevLett.69.2863}
\BIBentrySTDinterwordspacing

\bibitem{white_1993}
\BIBentryALTinterwordspacing
------, ``Density-matrix algorithms for quantum renormalization groups,''
  \emph{Physical Review B}, vol.~48, no.~14, pp. 10\,345--10\,356, Oct. 1993,
  \href{https://doi.org/10.1103/PhysRevB.48.10345}{doi:
  10.1103/PhysRevB.48.10345}. [Online]. Available:
  \url{https://link.aps.org/doi/10.1103/PhysRevB.48.10345}
\BIBentrySTDinterwordspacing

\bibitem{takasaki_1999}
\BIBentryALTinterwordspacing
H.~Takasaki, T.~Hikihara, and T.~Nishino, ``Fixed {{Point}} of the {{Finite
  System DMRG}},'' \emph{Journal of the Physical Society of Japan}, vol.~68,
  no.~5, pp. 1537--1540, May 1999,
  \href{https://doi.org/10.1143/JPSJ.68.1537}{doi: 10.1143/JPSJ.68.1537}.
  [Online]. Available: \url{https://journals.jps.jp/doi/10.1143/JPSJ.68.1537}
\BIBentrySTDinterwordspacing

\bibitem{schollwock_2005}
\BIBentryALTinterwordspacing
U.~Schollw{\"o}ck, ``The density-matrix renormalization group,'' \emph{Reviews
  of Modern Physics}, vol.~77, no.~1, pp. 259--315, Apr. 2005,
  \href{https://doi.org/10.1103/RevModPhys.77.259}{doi:
  10.1103/RevModPhys.77.259}. [Online]. Available:
  \url{https://link.aps.org/doi/10.1103/RevModPhys.77.259}
\BIBentrySTDinterwordspacing

\bibitem{schollwock_2011a}
\BIBentryALTinterwordspacing
------, ``The density-matrix renormalization group in the age of matrix product
  states,'' \emph{Annals of Physics}, vol. 326, no.~1, pp. 96--192, Jan. 2011,
  \href{https://doi.org/10.1016/j.aop.2010.09.012}{doi:
  10.1016/j.aop.2010.09.012}. [Online]. Available:
  \url{https://www.sciencedirect.com/science/article/pii/S0003491610001752}
\BIBentrySTDinterwordspacing

\bibitem{ostlund_1995}
\BIBentryALTinterwordspacing
S.~{\"O}stlund and S.~Rommer, ``Thermodynamic {{Limit}} of {{Density Matrix
  Renormalization}},'' \emph{Physical Review Letters}, vol.~75, no.~19, pp.
  3537--3540, Nov. 1995,
  \href{https://doi.org/10.1103/PhysRevLett.75.3537}{doi:
  10.1103/PhysRevLett.75.3537}. [Online]. Available:
  \url{https://link.aps.org/doi/10.1103/PhysRevLett.75.3537}
\BIBentrySTDinterwordspacing

\bibitem{dukelsky_1998}
\BIBentryALTinterwordspacing
J.~Dukelsky, M.~A. {Mart{\'i}n-Delgado}, T.~Nishino, and G.~Sierra,
  ``Equivalence of the variational matrix product method and the density matrix
  renormalization group applied to spin chains,'' \emph{Europhysics Letters},
  vol.~43, no.~4, p. 457, Aug. 1998,
  \href{https://doi.org/10.1209/epl/i1998-00381-x}{doi:
  10.1209/epl/i1998-00381-x}. [Online]. Available:
  \url{https://iopscience.iop.org/article/10.1209/epl/i1998-00381-x/meta}
\BIBentrySTDinterwordspacing

\bibitem{verstraete_2004}
\BIBentryALTinterwordspacing
F.~Verstraete, D.~Porras, and J.~I. Cirac, ``Density {{Matrix Renormalization
  Group}} and {{Periodic Boundary Conditions}}: {{A Quantum Information
  Perspective}},'' \emph{Physical Review Letters}, vol.~93, no.~22, p. 227205,
  Nov. 2004, \href{https://doi.org/10.1103/PhysRevLett.93.227205}{doi:
  10.1103/PhysRevLett.93.227205}. [Online]. Available:
  \url{https://link.aps.org/doi/10.1103/PhysRevLett.93.227205}
\BIBentrySTDinterwordspacing

\bibitem{beenakker_1991}
\BIBentryALTinterwordspacing
C.~W.~J. Beenakker, ``Theory of {{Coulomb-blockade}} oscillations in the
  conductance of a quantum dot,'' \emph{Physical Review B}, vol.~44, no.~4, pp.
  1646--1656, Jul. 1991, \href{https://doi.org/10.1103/PhysRevB.44.1646}{doi:
  10.1103/PhysRevB.44.1646}. [Online]. Available:
  \url{https://link.aps.org/doi/10.1103/PhysRevB.44.1646}
\BIBentrySTDinterwordspacing

\bibitem{livermore_1996}
\BIBentryALTinterwordspacing
C.~Livermore, C.~H. Crouch, R.~M. Westervelt, K.~L. Campman, and A.~C. Gossard,
  ``The {{Coulomb Blockade}} in {{Coupled Quantum Dots}},'' \emph{Science},
  vol. 274, no. 5291, pp. 1332--1335, Nov. 1996,
  \href{https://doi.org/10.1126/science.274.5291.1332}{doi:
  10.1126/science.274.5291.1332}. [Online]. Available:
  \url{https://www.science.org/doi/10.1126/science.274.5291.1332}
\BIBentrySTDinterwordspacing

\bibitem{ono_2002}
\BIBentryALTinterwordspacing
K.~Ono, D.~G. Austing, Y.~Tokura, and S.~Tarucha, ``Current {{Rectification}}
  by {{Pauli Exclusion}} in a {{Weakly Coupled Double Quantum Dot System}},''
  \emph{Science}, vol. 297, no. 5585, pp. 1313--1317, Aug. 2002,
  \href{https://doi.org/10.1126/science.1070958}{doi: 10.1126/science.1070958}.
  [Online]. Available:
  \url{https://www.science.org/doi/10.1126/science.1070958}
\BIBentrySTDinterwordspacing

\bibitem{elzerman_2003}
\BIBentryALTinterwordspacing
J.~M. Elzerman, R.~Hanson, J.~S. Greidanus, L.~H. {Willems van Beveren},
  S.~De~Franceschi, L.~M.~K. Vandersypen, S.~Tarucha, and L.~P. Kouwenhoven,
  ``Few-electron quantum dot circuit with integrated charge read out,''
  \emph{Physical Review B}, vol.~67, no.~16, p. 161308, Apr. 2003,
  \href{https://doi.org/10.1103/PhysRevB.67.161308}{doi:
  10.1103/PhysRevB.67.161308}. [Online]. Available:
  \url{https://link.aps.org/doi/10.1103/PhysRevB.67.161308}
\BIBentrySTDinterwordspacing

\bibitem{hofmann_1995}
\BIBentryALTinterwordspacing
F.~Hofmann, T.~Heinzel, D.~Wharam, J.~Kotthaus, G.~B{\"o}hm, W.~Klein,
  G.~Tr{\"a}nkle, and G.~Weimann, ``Single electron switching in a parallel
  quantum dot,'' \emph{Physical Review B}, vol.~51, no.~19, pp.
  13\,872--13\,875, May 1995,
  \href{https://doi.org/10.1103/PhysRevB.51.13872}{doi:
  10.1103/PhysRevB.51.13872}. [Online]. Available:
  \url{https://link.aps.org/doi/10.1103/PhysRevB.51.13872}
\BIBentrySTDinterwordspacing

\bibitem{kouwenhoven_1997}
\BIBentryALTinterwordspacing
L.~P. Kouwenhoven, C.~M. Marcus, P.~L. McEuen, S.~Tarucha, R.~M. Westervelt,
  and N.~S. Wingreen, ``Electron {{Transport}} in {{Quantum Dots}},'' in
  \emph{Mesoscopic {{Electron Transport}}}, L.~L. Sohn, L.~P. Kouwenhoven, and
  G.~Sch{\"o}n, Eds.\hskip 1em plus 0.5em minus 0.4em\relax {Dordrecht}:
  {Springer Netherlands}, 1997, pp. 105--214. [Online]. Available:
  \url{http://link.springer.com/10.1007/978-94-015-8839-3_4}
\BIBentrySTDinterwordspacing

\bibitem{schroer_2007}
\BIBentryALTinterwordspacing
D.~Schr{\"o}er, A.~D. Greentree, L.~Gaudreau, K.~Eberl, L.~C.~L. Hollenberg,
  J.~P. Kotthaus, and S.~Ludwig, ``Electrostatically defined serial triple
  quantum dot charged with few electrons,'' \emph{Physical Review B}, vol.~76,
  no.~7, p. 075306, Aug. 2007,
  \href{https://doi.org/10.1103/PhysRevB.76.075306}{doi:
  10.1103/PhysRevB.76.075306}. [Online]. Available:
  \url{https://link.aps.org/doi/10.1103/PhysRevB.76.075306}
\BIBentrySTDinterwordspacing

\bibitem{kalantre_2019}
\BIBentryALTinterwordspacing
S.~S. Kalantre, J.~P. Zwolak, S.~Ragole, X.~Wu, N.~M. Zimmerman, M.~D. Stewart,
  and J.~M. Taylor, ``Machine {{Learning}} techniques for state recognition and
  auto-tuning in quantum dots,'' \emph{npj Quantum Information}, vol.~5, no.~1,
  p.~6, Dec. 2019, \href{https://doi.org/10.1038/s41534-018-0118-7}{doi:
  10.1038/s41534-018-0118-7}. [Online]. Available:
  \url{http://arxiv.org/abs/1712.04914}
\BIBentrySTDinterwordspacing

\bibitem{march_1983}
\BIBentryALTinterwordspacing
N.~H. March, ``Origins\textemdash{{The Thomas-Fermi Theory}},'' in \emph{Theory
  of the {{Inhomogeneous Electron Gas}}}, ser. Physics of {{Solids}} and
  {{Liquids}}, S.~Lundqvist and N.~H. March, Eds.\hskip 1em plus 0.5em minus
  0.4em\relax {Boston, MA}: {Springer US}, 1983, pp. 1--77. [Online].
  Available: \url{https://doi.org/10.1007/978-1-4899-0415-7_1}
\BIBentrySTDinterwordspacing

\bibitem{merzbacher_1998a}
\BIBentryALTinterwordspacing
E.~Merzbacher, \emph{Quantum Mechanics}.\hskip 1em plus 0.5em minus 0.4em\relax
  {Wiley}, 1998. [Online]. Available:
  \url{https://books.google.de/books?id=6Ja_QgAACAAJ}
\BIBentrySTDinterwordspacing

\bibitem{dicarlo_2004}
\BIBentryALTinterwordspacing
L.~DiCarlo, H.~J. Lynch, A.~C. Johnson, L.~I. Childress, K.~Crockett, C.~M.
  Marcus, M.~P. Hanson, and A.~C. Gossard, ``Differential {{Charge Sensing}}
  and {{Charge Delocalization}} in a {{Tunable Double Quantum Dot}},''
  \emph{Physical Review Letters}, vol.~92, no.~22, p. 226801, Jun. 2004,
  \href{https://doi.org/10.1103/PhysRevLett.92.226801}{doi:
  10.1103/PhysRevLett.92.226801}. [Online]. Available:
  \url{https://link.aps.org/doi/10.1103/PhysRevLett.92.226801}
\BIBentrySTDinterwordspacing

\bibitem{petta_2004}
\BIBentryALTinterwordspacing
J.~R. Petta, A.~C. Johnson, C.~M. Marcus, M.~P. Hanson, and A.~C. Gossard,
  ``Manipulation of a {{Single Charge}} in a {{Double Quantum Dot}},''
  \emph{Physical Review Letters}, vol.~93, no.~18, p. 186802, Oct. 2004,
  \href{https://doi.org/10.1103/PhysRevLett.93.186802}{doi:
  10.1103/PhysRevLett.93.186802}. [Online]. Available:
  \url{https://link.aps.org/doi/10.1103/PhysRevLett.93.186802}
\BIBentrySTDinterwordspacing

\bibitem{hatano_2005}
\BIBentryALTinterwordspacing
T.~Hatano, M.~Stopa, and S.~Tarucha, ``Single-{{Electron Delocalization}} in
  {{Hybrid Vertical-Lateral Double Quantum Dots}},'' \emph{Science}, vol. 309,
  no. 5732, pp. 268--271, Jul. 2005,
  \href{https://doi.org/10.1126/science.1111205}{doi: 10.1126/science.1111205}.
  [Online]. Available:
  \url{https://www.science.org/doi/10.1126/science.1111205}
\BIBentrySTDinterwordspacing

\bibitem{huttel_2005}
\BIBentryALTinterwordspacing
A.~K. H{\"u}ttel, S.~Ludwig, H.~Lorenz, K.~Eberl, and J.~P. Kotthaus, ``Direct
  control of the tunnel splitting in a one-electron double quantum dot,''
  \emph{Physical Review B}, vol.~72, no.~8, p. 081310, Aug. 2005,
  \href{https://doi.org/10.1103/PhysRevB.72.081310}{doi:
  10.1103/PhysRevB.72.081310}. [Online]. Available:
  \url{https://link.aps.org/doi/10.1103/PhysRevB.72.081310}
\BIBentrySTDinterwordspacing

\bibitem{pioro-ladriere_2005}
\BIBentryALTinterwordspacing
M.~{Pioro-Ladri{\`e}re}, M.~R. Abolfath, P.~Zawadzki, J.~Lapointe, S.~A.
  Studenikin, A.~S. Sachrajda, and P.~Hawrylak, ``Charge sensing of an
  artificial \$\{\textbackslash
  mathrm\{\vphantom{\}\}}{{H}}\vphantom\{\}\vphantom\{\}\_\{2\}\^\{+\}\$
  molecule in lateral quantum dots,'' \emph{Physical Review B}, vol.~72,
  no.~12, p. 125307, Sep. 2005,
  \href{https://doi.org/10.1103/PhysRevB.72.125307}{doi:
  10.1103/PhysRevB.72.125307}. [Online]. Available:
  \url{https://link.aps.org/doi/10.1103/PhysRevB.72.125307}
\BIBentrySTDinterwordspacing

\bibitem{zhang_2006}
\BIBentryALTinterwordspacing
L.-X. Zhang, D.~V. Melnikov, and J.-P. Leburton, ``Exchange interaction and
  stability diagram of coupled quantum dots in magnetic fields,''
  \emph{Physical Review B}, vol.~74, no.~20, p. 205306, Nov. 2006,
  \href{https://doi.org/10.1103/PhysRevB.74.205306}{doi:
  10.1103/PhysRevB.74.205306}. [Online]. Available:
  \url{https://link.aps.org/doi/10.1103/PhysRevB.74.205306}
\BIBentrySTDinterwordspacing

\bibitem{stafford_1994}
\BIBentryALTinterwordspacing
C.~A. Stafford and S.~Das~Sarma, ``Collective {{Coulomb}} blockade in an array
  of quantum dots: {{A Mott-Hubbard}} approach,'' \emph{Physical Review
  Letters}, vol.~72, no.~22, pp. 3590--3593, May 1994,
  \href{https://doi.org/10.1103/PhysRevLett.72.3590}{doi:
  10.1103/PhysRevLett.72.3590}. [Online]. Available:
  \url{https://link.aps.org/doi/10.1103/PhysRevLett.72.3590}
\BIBentrySTDinterwordspacing

\bibitem{kotlyar_1998}
\BIBentryALTinterwordspacing
R.~Kotlyar, C.~A. Stafford, and S.~Das~Sarma, ``Correlated charge polarization
  in a chain of coupled quantum dots,'' \emph{Physical Review B}, vol.~58,
  no.~4, pp. R1746--R1749, Jul. 1998,
  \href{https://doi.org/10.1103/PhysRevB.58.R1746}{doi:
  10.1103/PhysRevB.58.R1746}. [Online]. Available:
  \url{https://link.aps.org/doi/10.1103/PhysRevB.58.R1746}
\BIBentrySTDinterwordspacing

\bibitem{jefferson_1996}
\BIBentryALTinterwordspacing
J.~H. Jefferson and W.~H{\"a}usler, ``Effective charge-spin models for quantum
  dots,'' \emph{Physical Review B}, vol.~54, no.~7, pp. 4936--4947, Aug. 1996,
  \href{https://doi.org/10.1103/PhysRevB.54.4936}{doi:
  10.1103/PhysRevB.54.4936}. [Online]. Available:
  \url{https://link.aps.org/doi/10.1103/PhysRevB.54.4936}
\BIBentrySTDinterwordspacing

\bibitem{gaudreau_2006}
\BIBentryALTinterwordspacing
L.~Gaudreau, S.~A. Studenikin, A.~S. Sachrajda, P.~Zawadzki, A.~Kam,
  J.~Lapointe, M.~Korkusinski, and P.~Hawrylak, ``Stability {{Diagram}} of a
  {{Few-Electron Triple Dot}},'' \emph{Physical Review Letters}, vol.~97,
  no.~3, p. 036807, Jul. 2006,
  \href{https://doi.org/10.1103/PhysRevLett.97.036807}{doi:
  10.1103/PhysRevLett.97.036807}. [Online]. Available:
  \url{https://link.aps.org/doi/10.1103/PhysRevLett.97.036807}
\BIBentrySTDinterwordspacing

\bibitem{korkusinski_2007}
\BIBentryALTinterwordspacing
M.~Korkusinski, I.~P. Gimenez, P.~Hawrylak, L.~Gaudreau, S.~A. Studenikin, and
  A.~S. Sachrajda, ``Topological {{Hunds}} rules and the electronic properties
  of a triple lateral quantum dot molecule,'' \emph{Physical Review B},
  vol.~75, no.~11, p. 115301, Mar. 2007,
  \href{https://doi.org/10.1103/PhysRevB.75.115301}{doi:
  10.1103/PhysRevB.75.115301}. [Online]. Available:
  \url{https://link.aps.org/doi/10.1103/PhysRevB.75.115301}
\BIBentrySTDinterwordspacing

\bibitem{yang_2011}
\BIBentryALTinterwordspacing
S.~Yang, X.~Wang, and S.~Das~Sarma, ``Generic {{Hubbard}} model description of
  semiconductor quantum-dot spin qubits,'' \emph{Physical Review B}, vol.~83,
  no.~16, p. 161301, Apr. 2011,
  \href{https://doi.org/10.1103/PhysRevB.83.161301}{doi:
  10.1103/PhysRevB.83.161301}. [Online]. Available:
  \url{https://link.aps.org/doi/10.1103/PhysRevB.83.161301}
\BIBentrySTDinterwordspacing

\bibitem{dassarma_2011}
\BIBentryALTinterwordspacing
S.~Das~Sarma, X.~Wang, and S.~Yang, ``Hubbard model description of silicon spin
  qubits: {{Charge}} stability diagram and tunnel coupling in {{Si}} double
  quantum dots,'' \emph{Physical Review B}, vol.~83, no.~23, p. 235314, Jun.
  2011, \href{https://doi.org/10.1103/PhysRevB.83.235314}{doi:
  10.1103/PhysRevB.83.235314}. [Online]. Available:
  \url{https://link.aps.org/doi/10.1103/PhysRevB.83.235314}
\BIBentrySTDinterwordspacing

\bibitem{wang_2011b}
\BIBentryALTinterwordspacing
X.~Wang, S.~Yang, and S.~Das~Sarma, ``Quantum theory of the charge-stability
  diagram of semiconductor double-quantum-dot systems,'' \emph{Physical Review
  B}, vol.~84, no.~11, p. 115301, Sep. 2011,
  \href{https://doi.org/10.1103/PhysRevB.84.115301}{doi:
  10.1103/PhysRevB.84.115301}. [Online]. Available:
  \url{https://link.aps.org/doi/10.1103/PhysRevB.84.115301}
\BIBentrySTDinterwordspacing

\bibitem{rassekh_2022}
\BIBentryALTinterwordspacing
A.~Rassekh, M.~Shalchian, J.-M. Sallese, and F.~Jazaeri, ``Tunneling {{Current
  Through}} a {{Double Quantum Dots System}},'' \emph{IEEE Access}, vol.~10,
  pp. 75\,245--75\,256, 2022,
  \href{https://doi.org/10.1109/ACCESS.2022.3190617}{doi:
  10.1109/ACCESS.2022.3190617}. [Online]. Available:
  \url{https://ieeexplore.ieee.org/document/9828406/}
\BIBentrySTDinterwordspacing

\bibitem{storcz_2005}
\BIBentryALTinterwordspacing
M.~J. Storcz, U.~Hartmann, S.~Kohler, and F.~K. Wilhelm, ``Intrinsic phonon
  decoherence and quantum gates in coupled lateral quantum-dot charge qubits,''
  \emph{Physical Review B}, vol.~72, no.~23, p. 235321, Dec. 2005,
  \href{https://doi.org/10.1103/PhysRevB.72.235321}{doi:
  10.1103/PhysRevB.72.235321}. [Online]. Available:
  \url{https://link.aps.org/doi/10.1103/PhysRevB.72.235321}
\BIBentrySTDinterwordspacing

\bibitem{stavrou_2005}
\BIBentryALTinterwordspacing
V.~N. Stavrou and X.~Hu, ``Charge decoherence in laterally coupled quantum dots
  due to electron-phonon interactions,'' \emph{Physical Review B}, vol.~72,
  no.~7, p. 075362, Aug. 2005,
  \href{https://doi.org/10.1103/PhysRevB.72.075362}{doi:
  10.1103/PhysRevB.72.075362}. [Online]. Available:
  \url{https://link.aps.org/doi/10.1103/PhysRevB.72.075362}
\BIBentrySTDinterwordspacing

\bibitem{witzel_2006}
\BIBentryALTinterwordspacing
W.~M. Witzel and S.~Das~Sarma, ``Quantum theory for electron spin decoherence
  induced by nuclear spin dynamics in semiconductor quantum computer
  architectures: {{Spectral}} diffusion of localized electron spins in the
  nuclear solid-state environment,'' \emph{Physical Review B}, vol.~74, no.~3,
  p. 035322, Jul. 2006, \href{https://doi.org/10.1103/PhysRevB.74.035322}{doi:
  10.1103/PhysRevB.74.035322}. [Online]. Available:
  \url{https://link.aps.org/doi/10.1103/PhysRevB.74.035322}
\BIBentrySTDinterwordspacing

\bibitem{taylor_2007}
\BIBentryALTinterwordspacing
J.~M. Taylor, J.~R. Petta, A.~C. Johnson, A.~Yacoby, C.~M. Marcus, and M.~D.
  Lukin, ``Relaxation, dephasing, and quantum control of electron spins in
  double quantum dots,'' \emph{Physical Review B}, vol.~76, no.~3, p. 035315,
  Jul. 2007, \href{https://doi.org/10.1103/PhysRevB.76.035315}{doi:
  10.1103/PhysRevB.76.035315}. [Online]. Available:
  \url{https://link.aps.org/doi/10.1103/PhysRevB.76.035315}
\BIBentrySTDinterwordspacing

\bibitem{gimenez_2009}
\BIBentryALTinterwordspacing
I.~P. Gimenez, C.-Y. Hsieh, M.~Korkusinski, and P.~Hawrylak,
  ``Charged-impurity-induced dephasing of a voltage-controlled coded qubit
  based on electron spin in a triple quantum dot,'' \emph{Physical Review B},
  vol.~79, no.~20, p. 205311, May 2009,
  \href{https://doi.org/10.1103/PhysRevB.79.205311}{doi:
  10.1103/PhysRevB.79.205311}. [Online]. Available:
  \url{https://link.aps.org/doi/10.1103/PhysRevB.79.205311}
\BIBentrySTDinterwordspacing

\bibitem{nguyen_2011a}
\BIBentryALTinterwordspacing
N.~T.~T. Nguyen and S.~Das~Sarma, ``Impurity effects on semiconductor quantum
  bits in coupled quantum dots,'' \emph{Physical Review B}, vol.~83, no.~23, p.
  235322, Jun. 2011, \href{https://doi.org/10.1103/PhysRevB.83.235322}{doi:
  10.1103/PhysRevB.83.235322}. [Online]. Available:
  \url{https://link.aps.org/doi/10.1103/PhysRevB.83.235322}
\BIBentrySTDinterwordspacing

\bibitem{lennon_2019}
\BIBentryALTinterwordspacing
D.~T. Lennon, H.~Moon, L.~C. Camenzind, L.~Yu, D.~M. Zumb{\"u}hl, G.~a.~D.
  Briggs, M.~A. Osborne, E.~A. Laird, and N.~Ares, ``Efficiently measuring a
  quantum device using machine learning,'' \emph{npj Quantum Information},
  vol.~5, no.~1, pp. 1--8, Sep. 2019,
  \href{https://doi.org/10.1038/s41534-019-0193-4}{doi:
  10.1038/s41534-019-0193-4}. [Online]. Available:
  \url{https://www.nature.com/articles/s41534-019-0193-4}
\BIBentrySTDinterwordspacing

\bibitem{oakes_2021}
\BIBentryALTinterwordspacing
G.~A. Oakes, J.~Duan, J.~J.~L. Morton, A.~Lee, C.~G. Smith, and M.~F.~G. Zalba,
  ``Automatic virtual voltage extraction of a 2x2 array of quantum dots with
  machine learning,'' \emph{arXiv:2012.03685 [cond-mat, physics:quant-ph]}, May
  2021. [Online]. Available: \url{http://arxiv.org/abs/2012.03685}
\BIBentrySTDinterwordspacing

\bibitem{krause_2022}
\BIBentryALTinterwordspacing
O.~Krause, A.~Chatterjee, F.~Kuemmeth, and E.~Van~Nieuwenburg, ``Learning
  {{Coulomb Diamonds}} in {{Large Quantum Dot Arrays}},'' \emph{SciPost
  Physics}, vol.~13, no.~4, p. 084, Oct. 2022,
  \href{https://doi.org/10.21468/SciPostPhys.13.4.084}{doi:
  10.21468/SciPostPhys.13.4.084}. [Online]. Available:
  \url{https://scipost.org/10.21468/SciPostPhys.13.4.084}
\BIBentrySTDinterwordspacing

\bibitem{li_2023}
\BIBentryALTinterwordspacing
W.~Li, Z.~Liu, J.~Mu, Y.~Luo, D.~Pan, J.~Zhao, and H.~Q. Xu, ``Charge
  {{States}}, {{Triple Points}}, and {{Quadruple Points}} in an {{InAs Nanowire
  Triple Quantum Dot Revealed}} by an {{Integrated Charge Sensor}},''
  \emph{Advanced Quantum Technologies}, vol.~6, no.~5, p. 2200158, 2023,
  \href{https://doi.org/10.1002/qute.202200158}{doi: 10.1002/qute.202200158}.
  [Online]. Available:
  \url{https://onlinelibrary.wiley.com/doi/abs/10.1002/qute.202200158}
\BIBentrySTDinterwordspacing

\bibitem{darulova_2020a}
\BIBentryALTinterwordspacing
J.~Darulova, M.~Troyer, and M.~C. Cassidy, ``Evaluation of synthetic and
  experimental training data in supervised machine learning applied to charge
  state detection of quantum dots,'' \emph{arXiv:2005.08131 [cond-mat,
  physics:quant-ph]}, May 2020. [Online]. Available:
  \url{http://arxiv.org/abs/2005.08131}
\BIBentrySTDinterwordspacing

\bibitem{zwolak_2018}
\BIBentryALTinterwordspacing
J.~P. Zwolak, S.~S. Kalantre, X.~Wu, S.~Ragole, and J.~M. Taylor, ``{{QFlow}}
  lite dataset: {{A}} machine-learning approach to the charge states in quantum
  dot experiments,'' \emph{PLOS ONE}, vol.~13, no.~10, p. e0205844, Oct. 2018,
  \href{https://doi.org/10.1371/journal.pone.0205844}{doi:
  10.1371/journal.pone.0205844}. [Online]. Available:
  \url{https://journals.plos.org/plosone/article?id=10.1371/journal.pone.0205844}
\BIBentrySTDinterwordspacing

\bibitem{ziegler_2022}
\BIBentryALTinterwordspacing
J.~Ziegler, T.~McJunkin, E.~Joseph, S.~S. Kalantre, B.~Harpt, D.~Savage,
  M.~Lagally, M.~Eriksson, J.~M. Taylor, and J.~P. Zwolak, ``Toward {Robust}
  {Autotuning} of {Noisy} {Quantum} dot {Devices},'' \emph{Physical Review
  Applied}, vol.~17, no.~2, p. 024069, Feb. 2022, publisher: American Physical
  Society. [Online]. Available:
  \url{https://link.aps.org/doi/10.1103/PhysRevApplied.17.024069}
\BIBentrySTDinterwordspacing

\bibitem{johansson_2012}
\BIBentryALTinterwordspacing
J.~R. Johansson, P.~D. Nation, and F.~Nori, ``{{QuTiP}}: {{An}} open-source
  {{Python}} framework for the dynamics of open quantum systems,''
  \emph{Computer Physics Communications}, vol. 183, no.~8, pp. 1760--1772, Aug.
  2012, \href{https://doi.org/10.1016/j.cpc.2012.02.021}{doi:
  10.1016/j.cpc.2012.02.021}. [Online]. Available:
  \url{https://www.sciencedirect.com/science/article/pii/S0010465512000835}
\BIBentrySTDinterwordspacing

\bibitem{johansson_2013}
\BIBentryALTinterwordspacing
------, ``{{QuTiP}} 2: {{A Python}} framework for the dynamics of open quantum
  systems,'' \emph{Computer Physics Communications}, vol. 184, no.~4, pp.
  1234--1240, Apr. 2013, \href{https://doi.org/10.1016/j.cpc.2012.11.019}{doi:
  10.1016/j.cpc.2012.11.019}. [Online]. Available:
  \url{https://www.sciencedirect.com/science/article/pii/S0010465512003955}
\BIBentrySTDinterwordspacing

\bibitem{smidstrup_2019}
\BIBentryALTinterwordspacing
S.~Smidstrup, T.~Markussen, P.~Vancraeyveld, J.~Wellendorff, J.~Schneider,
  T.~Gunst, B.~Verstichel, D.~Stradi, P.~A. Khomyakov, U.~G. {Vej-Hansen},
  M.-E. Lee, S.~T. Chill, F.~Rasmussen, G.~Penazzi, F.~Corsetti,
  A.~Ojanper{\"a}, K.~Jensen, M.~L.~N. Palsgaard, U.~Martinez, A.~Blom,
  M.~Brandbyge, and K.~Stokbro, ``{{QuantumATK}}: An integrated platform of
  electronic and atomic-scale modelling tools,'' \emph{Journal of Physics:
  Condensed Matter}, vol.~32, no.~1, p. 015901, Oct. 2019,
  \href{https://doi.org/10.1088/1361-648X/ab4007}{doi:
  10.1088/1361-648X/ab4007}. [Online]. Available:
  \url{https://dx.doi.org/10.1088/1361-648X/ab4007}
\BIBentrySTDinterwordspacing

\bibitem{groth_2014}
\BIBentryALTinterwordspacing
C.~W. Groth, M.~Wimmer, A.~R. Akhmerov, and X.~Waintal, ``Kwant: A software
  package for quantum transport,'' \emph{New Journal of Physics}, vol.~16,
  no.~6, p. 063065, Jun. 2014,
  \href{https://doi.org/10.1088/1367-2630/16/6/063065}{doi:
  10.1088/1367-2630/16/6/063065}. [Online]. Available:
  \url{https://iopscience.iop.org/article/10.1088/1367-2630/16/6/063065}
\BIBentrySTDinterwordspacing

\bibitem{birner_2007}
S.~Birner, T.~Zibold, T.~Andlauer, T.~Kubis, M.~Sabathil, A.~Trellakis, and
  P.~Vogl, ``Nextnano: {{General Purpose}} 3-{{D Simulations}},'' \emph{IEEE
  Transactions on Electron Devices}, vol.~54, no.~9, pp. 2137--2142, Sep. 2007,
  \href{https://doi.org/10.1109/TED.2007.902871}{doi: 10.1109/TED.2007.902871}.

\bibitem{gao_2013a}
\BIBentryALTinterwordspacing
X.~Gao, E.~Nielsen, R.~P. Muller, R.~W. Young, A.~G. Salinger, N.~C. Bishop,
  M.~P. Lilly, and M.~S. Carroll, ``{Quantum computer aided design simulation
  and optimization of semiconductor quantum dots},'' \emph{Journal of Applied
  Physics}, vol. 114, no.~16, p. 164302, Oct. 2013,
  \href{https://doi.org/10.1063/1.4825209}{doi: 10.1063/1.4825209}. [Online].
  Available: \url{http://arxiv.org/abs/1403.7561}
\BIBentrySTDinterwordspacing

\bibitem{nielsen_2013}
\BIBentryALTinterwordspacing
E.~Nielsen, X.~Gao, I.~Kalashnikova, R.~P. Muller, A.~G. Salinger, and R.~W.
  Young, ``Qcad simulation and optimization of semiconductor double quantum
  dots,'' 12 2013. [Online]. Available:
  \url{https://www.osti.gov/biblio/1204068}
\BIBentrySTDinterwordspacing

\bibitem{kirsanskas_2017}
\BIBentryALTinterwordspacing
G.~Kir{\v s}anskas, J.~N. Pedersen, O.~Karlstr{\"o}m, M.~Leijnse, and
  A.~Wacker, ``{{QmeQ}} 1.0: {{An}} open-source {{Python}} package for
  calculations of transport through quantum dot devices,'' \emph{Computer
  Physics Communications}, vol. 221, pp. 317--342, Dec. 2017,
  \href{https://doi.org/10.1016/j.cpc.2017.07.024}{doi:
  10.1016/j.cpc.2017.07.024}. [Online]. Available:
  \url{https://www.sciencedirect.com/science/article/pii/S0010465517302515}
\BIBentrySTDinterwordspacing

\bibitem{ikegami_2019}
\BIBentryALTinterwordspacing
T.~Ikegami, K.~Fukuda, J.~Hattori, H.~Asai, and H.~Ota, ``A {{TCAD}} device
  simulator for exotic materials and its application to a negative-capacitance
  {{FET}},'' \emph{Journal of Computational Electronics}, vol.~18, no.~2, pp.
  534--542, Jun. 2019, \href{https://doi.org/10.1007/s10825-019-01313-7}{doi:
  10.1007/s10825-019-01313-7}. [Online]. Available:
  \url{https://doi.org/10.1007/s10825-019-01313-7}
\BIBentrySTDinterwordspacing

\bibitem{asai_2021a}
H.~Asai, S.~Iizuka, T.~Ikegami, J.~Hattori, K.~Fukuda, H.~Oka, K.~Kato, H.~Ota,
  and T.~Mori, ``Development of {{Integrated Device Simulator}} for {{Quantum
  Bit Design}}: {{Self-consistent Calculation}} for {{Quantum Transport}} and
  {{Qubit Operation}},'' in \emph{2021 5th {{IEEE Electron Devices Technology}}
  \& {{Manufacturing Conference}} ({{EDTM}})}, Apr. 2021, pp. 1--3,
  \href{https://doi.org/10.1109/EDTM50988.2021.9420978}{doi:
  10.1109/EDTM50988.2021.9420978}.

\bibitem{beaudoin_2022}
\BIBentryALTinterwordspacing
F.~Beaudoin, P.~Philippopoulos, C.~Zhou, I.~Kriekouki, M.~{Pioro-Ladri{\`e}re},
  H.~Guo, and P.~Galy, ``Robust technology computer-aided design of gated
  quantum dots at cryogenic temperature,'' \emph{Applied Physics Letters}, vol.
  120, no.~26, p. 264001, Jun. 2022,
  \href{https://doi.org/10.1063/5.0097202}{doi: 10.1063/5.0097202}. [Online].
  Available: \url{https://doi.org/10.1063/5.0097202}
\BIBentrySTDinterwordspacing

\bibitem{wasshuber_1997a}
C.~Wasshuber, H.~Kosina, and S.~Selberherr, ``{{SIMON-A}} simulator for
  single-electron tunnel devices and circuits,'' \emph{IEEE Transactions on
  Computer-Aided Design of Integrated Circuits and Systems}, vol.~16, no.~9,
  pp. 937--944, Sep. 1997, \href{https://doi.org/10.1109/43.658562}{doi:
  10.1109/43.658562}.

\bibitem{eendebak_2023}
\BIBentryALTinterwordspacing
P.~Eendebak, ``Qtt {{Documentation}},'' QuTech, {Delft}, Jan. 2023. [Online].
  Available: \url{https://qtt.readthedocs.io/_/downloads/en/dev/pdf}
\BIBentrySTDinterwordspacing

\bibitem{qudipy}
\BIBentryALTinterwordspacing
{Khromets, Bohdan}, ``Heterostructure development and quantum control for
  semiconductor qubits,'' Master's thesis, 2022. [Online]. Available:
  \url{http://hdl.handle.net/10012/17823}
\BIBentrySTDinterwordspacing

\bibitem{fleitmann_2023}
\BIBentryALTinterwordspacing
S.~Fleitmann, ``{C}haracterization of {D}istortions in {C}harge {S}tability
  {D}iagrams and their {S}imulation in {M}odeled {D}ata,'' Forschungszentrum
  Jülich GmbH Zentralbibliothek, Verlag, Tech. Rep., 2023. [Online].
  Available: \url{https://juser.fz-juelich.de/record/1019873}
\BIBentrySTDinterwordspacing

\bibitem{hader_paper}
F.~Hader, J.~Vogelbruch, S.~Humpohl, T.~Hangleiter, C.~Eguzo, S.~Heinen,
  S.~Meyer, and S.~van Waasen, ``On noise-sensitive automatic tuning of
  gate-defined sensor dots,'' \emph{IEEE Transactions on Quantum Engineering},
  vol.~4, pp. 1--18, 2023.

\bibitem{botzem_tuning_2018}
\BIBentryALTinterwordspacing
T.~Botzem, M.~D. Shulman, S.~Foletti, S.~P. Harvey, O.~E. Dial, P.~Bethke,
  P.~Cerfontaine, R.~P.~G. McNeil, D.~Mahalu, V.~Umansky, A.~Ludwig, A.~Wieck,
  D.~Schuh, D.~Bougeard, A.~Yacoby, and H.~Bluhm,
  ``\BIBforeignlanguage{en}{Tuning {Methods} for {Semiconductor} {Spin}
  {Qubits}},'' \emph{\BIBforeignlanguage{en}{Physical Review Applied}},
  vol.~10, no.~5, p. 054026, Nov. 2018. [Online]. Available:
  \url{https://link.aps.org/doi/10.1103/PhysRevApplied.10.054026}
\BIBentrySTDinterwordspacing

\bibitem{maradan_gaas_2014}
\BIBentryALTinterwordspacing
D.~Maradan, L.~Casparis, T.-M. Liu, D.~E.~F. Biesinger, C.~P. Scheller, D.~M.
  Zumbühl, J.~D. Zimmerman, and A.~C. Gossard,
  ``\BIBforeignlanguage{en}{{GaAs} {Quantum} {Dot} {Thermometry} {Using}
  {Direct} {Transport} and {Charge} {Sensing}},''
  \emph{\BIBforeignlanguage{en}{Journal of Low Temperature Physics}}, vol. 175,
  no.~5, pp. 784--798, Jun. 2014. [Online]. Available:
  \url{https://doi.org/10.1007/s10909-014-1169-6}
\BIBentrySTDinterwordspacing

\bibitem{electronicNoise}
\BIBentryALTinterwordspacing
S.~Yuan, T.~Gao, W.~Cao, Z.~Pan, J.~Liu, J.~Shi, and W.~Hong,
  ``\BIBforeignlanguage{en}{The {Characterization} of {Electronic} {Noise} in
  the {Charge} {Transport} through {Single}-{Molecule} {Junctions}},''
  \emph{\BIBforeignlanguage{en}{Small Methods}}, vol.~5, no.~3, p. 2001064,
  Mar. 2021. [Online]. Available:
  \url{https://onlinelibrary.wiley.com/doi/10.1002/smtd.202001064}
\BIBentrySTDinterwordspacing

\bibitem{pink_noise}
\BIBentryALTinterwordspacing
J.~Timmer and M.~Koenig, ``On generating power law noise.'' \emph{Astronomy and
  Astrophysics}, vol. 300, p. 707, Aug. 1995. [Online]. Available:
  \url{https://ui.adsabs.harvard.edu/abs/1995A&A...300..707T}
\BIBentrySTDinterwordspacing

\bibitem{colorednoise}
\BIBentryALTinterwordspacing
F.~Patzelt, ``colorednoise.py,'' 2017. [Online]. Available:
  \url{https://github.com/felixpatzelt/colorednoise}
\BIBentrySTDinterwordspacing

\bibitem{rtn}
\BIBentryALTinterwordspacing
X.~Chen, L.~Wang, B.~Li, Y.~Wang, X.~Li, Y.~Liu, and H.~Yang, ``Modeling
  {Random} {Telegraph} {Noise} as a {Randomness} {Source} and its {Application}
  in {True} {Random} {Number} {Generation},'' \emph{IEEE Transactions on
  Computer-Aided Design of Integrated Circuits and Systems}, vol.~35, no.~9,
  pp. 1435--1448, Sep. 2016. [Online]. Available:
  \url{http://ieeexplore.ieee.org/document/7362179/}
\BIBentrySTDinterwordspacing

\bibitem{thermal_noise}
\BIBentryALTinterwordspacing
J.~B. Johnson, ``Thermal agitation of electricity in conductors,'' \emph{Phys.
  Rev.}, vol.~32, pp. 97--109, Jul 1928. [Online]. Available:
  \url{https://link.aps.org/doi/10.1103/PhysRev.32.97}
\BIBentrySTDinterwordspacing

\bibitem{shot_noise}
\BIBentryALTinterwordspacing
W.~Schottky, ``{Über spontane Stromschwankungen in verschiedenen
  Elektrizitätsleitern},'' Jan. 1918. [Online]. Available:
  \url{https://doi.org/10.1002/andp.19183622304}
\BIBentrySTDinterwordspacing

\bibitem{simcats_github}
\BIBentryALTinterwordspacing
F.~Hader, ``{SimCATS},'' Dec. 2023, original-date: 2023-11-20T12:28:44Z.
  [Online]. Available: \url{https://github.com/f-hader/SimCATS}
\BIBentrySTDinterwordspacing

\bibitem{volk_loading_2019}
\BIBentryALTinterwordspacing
C.~Volk, A.~M.~J. Zwerver, U.~Mukhopadhyay, P.~T. Eendebak, C.~J. van Diepen,
  J.~P. Dehollain, T.~Hensgens, T.~Fujita, C.~Reichl, W.~Wegscheider, and
  L.~M.~K. Vandersypen, ``\BIBforeignlanguage{en}{Loading a quantum-dot based
  “{Qubyte}” register},'' \emph{\BIBforeignlanguage{en}{npj Quantum
  Information}}, vol.~5, no.~1, pp. 1--8, Apr. 2019, number: 1 Publisher:
  Nature Publishing Group. [Online]. Available:
  \url{https://www.nature.com/articles/s41534-019-0146-y}
\BIBentrySTDinterwordspacing

\bibitem{welch}
P.~Welch, ``The use of fast fourier transform for the estimation of power
  spectra: A method based on time averaging over short, modified
  periodograms,'' \emph{IEEE Transactions on Audio and Electroacoustics},
  vol.~15, no.~2, pp. 70--73, 1967.

\bibitem{scipy}
P.~Virtanen, R.~Gommers, T.~E. Oliphant, M.~Haberland, T.~Reddy, D.~Cournapeau,
  E.~Burovski, P.~Peterson, W.~Weckesser, J.~Bright, S.~J. {van der Walt},
  M.~Brett, J.~Wilson, K.~J. Millman, N.~Mayorov, A.~R.~J. Nelson, E.~Jones,
  R.~Kern, E.~Larson, C.~J. Carey, {\.I}.~Polat, Y.~Feng, E.~W. Moore,
  J.~{VanderPlas}, D.~Laxalde, J.~Perktold, R.~Cimrman, I.~Henriksen, E.~A.
  Quintero, C.~R. Harris, A.~M. Archibald, A.~H. Ribeiro, F.~Pedregosa, P.~{van
  Mulbregt}, and {SciPy 1.0 Contributors}, ``{{SciPy} 1.0: Fundamental
  Algorithms for Scientific Computing in Python},'' \emph{Nature Methods},
  vol.~17, pp. 261--272, 2020.

\bibitem{inception_score}
\BIBentryALTinterwordspacing
T.~Salimans, I.~Goodfellow, W.~Zaremba, V.~Cheung, A.~Radford, and X.~Chen,
  ``Improved {Techniques} for {Training} {GANs},'' Jun. 2016. [Online].
  Available: \url{http://arxiv.org/abs/1606.03498}
\BIBentrySTDinterwordspacing

\bibitem{frechet_distance}
M.~Heusel, H.~Ramsauer, T.~Unterthiner, B.~Nessler, and S.~Hochreiter, ``Gans
  trained by a two time-scale update rule converge to a local nash
  equilibrium,'' in \emph{Proceedings of the 31st International Conference on
  Neural Information Processing Systems}, ser. NIPS'17.\hskip 1em plus 0.5em
  minus 0.4em\relax Red Hook, NY, USA: Curran Associates Inc., 2017, p.
  6629–6640.

\bibitem{precision_recall}
\BIBentryALTinterwordspacing
A.~M. Alaa, B.~van Breugel, E.~Saveliev, and M.~van~der Schaar, ``How
  {Faithful} is your {Synthetic} {Data}? {Sample}-level {Metrics} for
  {Evaluating} and {Auditing} {Generative} {Models},'' \emph{arXiv:2102.08921
  [cs, stat]}, Feb. 2021. [Online]. Available:
  \url{http://arxiv.org/abs/2102.08921}
\BIBentrySTDinterwordspacing

\bibitem{deepSVDD_implementation}
\BIBentryALTinterwordspacing
L.~Ruff, ``Pytorch implementation of deep svdd,'' 2018. [Online]. Available:
  \url{https://github.com/lukasruff/Deep-SVDD-PyTorch}
\BIBentrySTDinterwordspacing

\bibitem{deepSVDD_paper}
\BIBentryALTinterwordspacing
L.~Ruff, R.~Vandermeulen, N.~Goernitz, L.~Deecke, S.~A. Siddiqui, A.~Binder,
  E.~M{\"u}ller, and M.~Kloft, ``Deep one-class classification,'' in
  \emph{Proceedings of the 35th International Conference on Machine Learning},
  ser. Proceedings of Machine Learning Research, J.~Dy and A.~Krause, Eds.,
  vol.~80.\hskip 1em plus 0.5em minus 0.4em\relax PMLR, 10--15 Jul 2018, pp.
  4393--4402. [Online]. Available:
  \url{https://proceedings.mlr.press/v80/ruff18a.html}
\BIBentrySTDinterwordspacing

\bibitem{albumentations}
A.~Buslaev, A.~Parinov, E.~Khvedchenya, V.~I. Iglovikov, and A.~A. Kalinin,
  ``{Albumentations: fast and flexible image augmentations},'' \emph{ArXiv
  e-prints}, 2018.

\bibitem{mnist}
L.~Deng, ``The mnist database of handwritten digit images for machine learning
  research [best of the web],'' \emph{IEEE Signal Processing Magazine},
  vol.~29, no.~6, pp. 141--142, 2012.

\end{thebibliography}

    \begin{IEEEbiography}[{\includegraphics[width=1in,height=1.25in,clip,keepaspectratio]{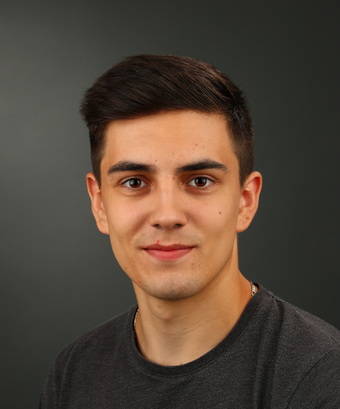}}]{Fabian Hader} received the B.Sc. degree in scientific programming and the M.Sc. degree in energy economics {\&} informatics from FH Aachen -  University of applied sciences, Jülich, Germany, in 2019 and 2021, respectively. He is currently pursuing a Ph.D. degree in engineering at University Duisburg-Essen, Duisburg/Essen, Germany.
    
    From 2019 to 2021, he was a Software Engineer at the Central Institute of Engineering, Electronics, and Analytics - Electronic Systems, Forschungszentrum Jülich GmbH, Jülich, Germany. His research interest focuses on the automatic tuning of \acp{qd}.
    \end{IEEEbiography}
    
    \begin{IEEEbiography}[{\includegraphics[width=1in,height=1.25in,clip,keepaspectratio]{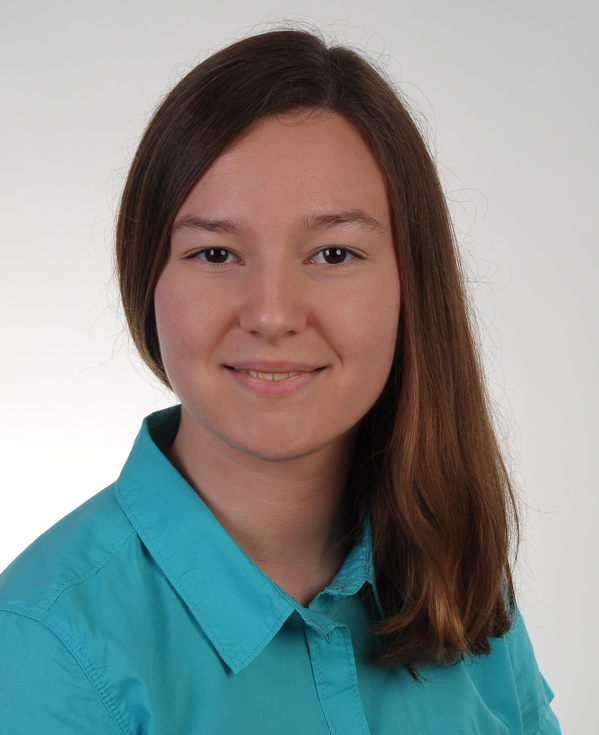}}]
    {Sarah Fleitmann} received the B.Sc.\ degree in scientific programming and the M.Sc.\ degree in applied mathematics and informatics from the FH Aachen -- University of Applied Sciences, Campus Jülich, Germany in 2020 and 2022, respectively.\\
    Since 2017, she works as a software engineer at the Central Institute of Engineering, Electronics and Analytics -- Electronic Systems at Forschungszentrum Jülich GmbH, Jülich, Germany. Her research interests include the automatic tuning of quantum dots for their operation as qubits.
    \end{IEEEbiography}
    
    \begin{IEEEbiography}[{\includegraphics[width=1in,height=1.25in,clip,keepaspectratio]{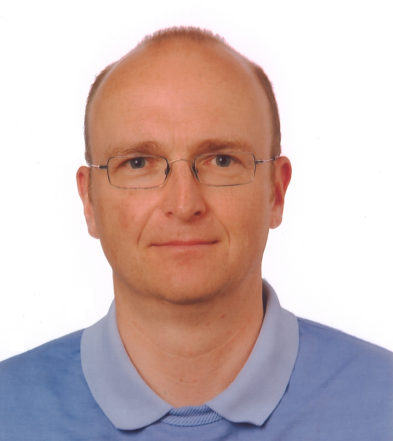}}]{Jan Vogelbruch} received the Dipl.Ing. and Dr.-Ing. degrees in electrical engineering from the RWTH Aachen University, Germany, in 1994 and 2003, respectively.\\
    In 1995, he joined Parsytec Computer GmbH, Aachen, Germany, as technical project manager for European cooperations. His focus has been on high-performance computing and image processing solutions, where he has been the technical leader for the company's part in several EC-funded projects.
    Since late 1998, he has been with the Central Institute of Engineering, Electronics, and Analytics - Electronic Systems at Forschungszentrum Jülich GmbH, Jülich, Germany. His research interests include parallel computing, signal and 3D image processing, fast reconstruction methods for high-resolution computer tomography, and automated defect detection. His current research focus is on the automatic tuning of semiconductor quantum dots.
    \end{IEEEbiography}
    
    \begin{IEEEbiography}[{\includegraphics[width=1in,height=1.25in,clip,keepaspectratio]{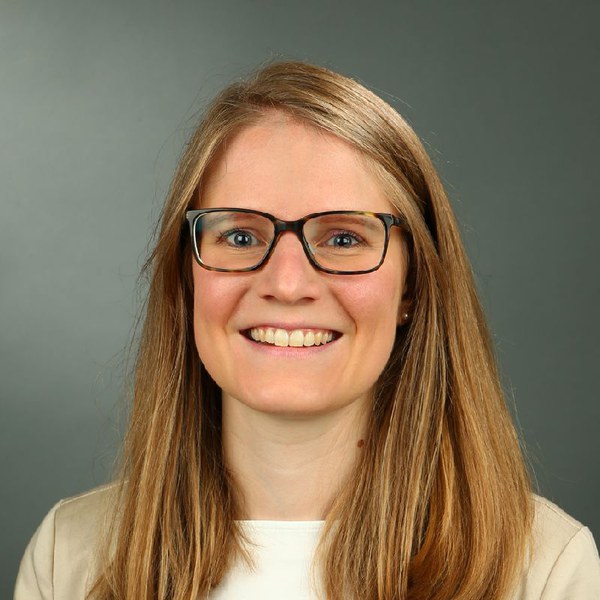}}]{Lotte Geck} received the B.Sc. and M.Sc. degrees from the RWTH Aachen University in 2013 and 2015 respectively. In 2016 she joined the Central Institute of Engineering, Electronics and Analytics – Electronic Systems at Forschungszentrum Jülich GmbH, Jülich, Germany. She received the Dr.-Ing. degree in 2021. Since 2022 she is a Junior Professor at Forschungszentrum Jülich and RWTH Aachen University. Her research interests include scalable electronic system solutions for quantum computing.
    \end{IEEEbiography}

    \begin{IEEEbiography}[{\includegraphics[width=1in,height=1.25in,clip,keepaspectratio]{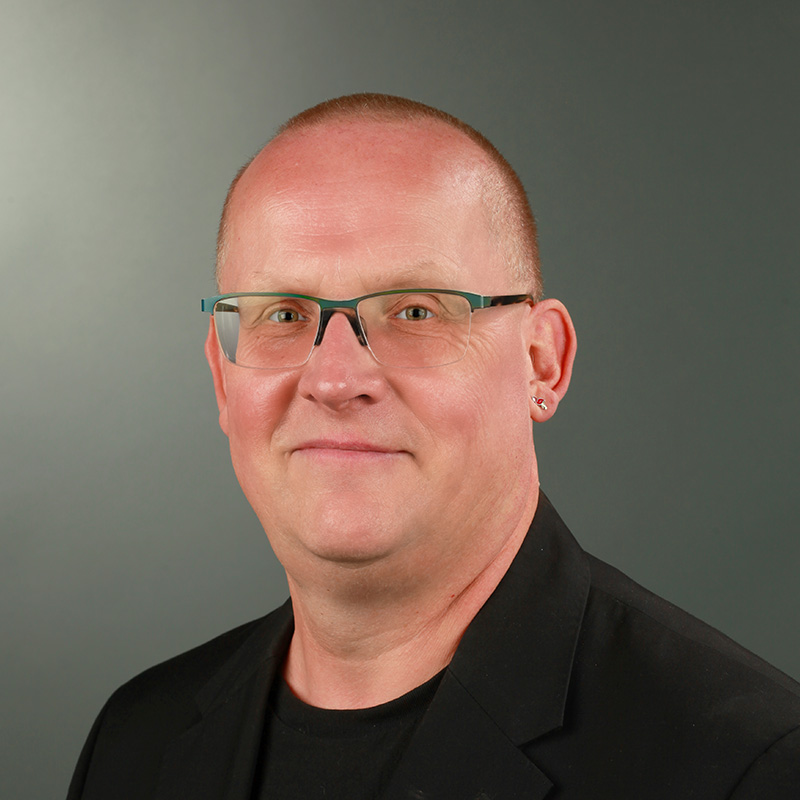}}]{Stefan van Waasen} received his diploma and doctor's degrees in electrical engineering from Gerhard-Mercator University, Duisburg, Germany, in 1994 and 1999, respectively. The topic of his doctoral thesis was optical receivers up to 60 Gb/s based on traveling wave amplifiers.\\
    In 1998, he joined Siemens Semiconductors/Infineon Technologies AG, Düsseldorf, Germany. His responsibility was BiCMOS and CMOS RF system development for highly integrated cordless systems like DECT and Bluetooth. In 2001, he changed into the IC development of front-end systems for high data rate optical communication systems. From 2004 to 2006, he was with the Stockholm Design Center responsible for the short-range analog, mixed-signal, and RF development for SoC CMOS solutions. From 2006 to 2010, he was responsible for the wireless RF system engineering in the area of SoC CMOS products at the headquarters in Munich, Germany, and later in the Design Center Duisburg, Duisburg. Since 2010, he has been the Director of the Central Institute of Engineering, Electronics, and Analytics - Electronic Systems at Forschungszentrum Jülich GmbH, Jülich, Germany. In 2014, he became a professor for measurement and sensor systems at the Communication Systems Chair of the University of Duisburg-Essen. His research is in the direction of complex measurement and detector systems, particularly on electronic systems for Quantum Computing.
    \end{IEEEbiography}

    \clearpage{}
    
    \setcounter{figure}{0} \renewcommand{\thefigure}{A.\arabic{figure}}
    \setcounter{table}{0} \renewcommand{\thetable}{A.\arabic{table}}

    \EOD

\end{document}